\documentclass[twoside,twocolumn,9pt]{article}

\usepackage{extsizes}
\usepackage[super,sort&compress,comma]{natbib} 
\usepackage[version=3]{mhchem}
\usepackage[left=1.5cm, right=1.5cm, top=1.785cm, bottom=2.0cm]{geometry}
\usepackage{balance}
\usepackage{mathptmx}
\usepackage{sectsty}
\usepackage{graphicx} 
\usepackage{lastpage}
\usepackage[format=plain,justification=justified,singlelinecheck=false,font={stretch=1.125,small,sf},labelfont=bf,labelsep=space]{caption}
\usepackage{float}
\usepackage{fancyhdr}
\usepackage{fnpos}
\usepackage[english]{babel}
\addto{\captionsenglish}{%
  
}
\usepackage{array}
\usepackage{droidsans}
\usepackage{charter}
\usepackage[T1]{fontenc}
\usepackage[usenames,dvipsnames]{xcolor}
\usepackage{setspace}
\usepackage[compact]{titlesec}
\usepackage{bm} 
\usepackage{float} 
\usepackage{bbm} 
\usepackage{enumitem,xcolor}
\usepackage{color}
\usepackage{amsfonts}
\usepackage{lmodern} 
\usepackage{comment}
\usepackage{amsmath,amssymb}
\usepackage{natbib}
\usepackage{ulem}
\usepackage{multirow}
\usepackage{array}

\usepackage{graphicx}
\usepackage{multirow}
\usepackage{array}
\usepackage{subcaption} 
\usepackage{geometry}
\usepackage{tikz} 
\usetikzlibrary{shapes,arrows,calc}
\usetikzlibrary{datavisualization}
\usetikzlibrary{datavisualization.formats.functions}
\usepackage{pgfplots} 
\usepackage{tikz}
\usetikzlibrary{matrix}

\newcommand\bi{\begin{itemize}}
\newcommand\ei{\end{itemize}}

\usepackage[colorlinks=true,citecolor=blue]{hyperref}
\hypersetup{colorlinks=true,citecolor=blue,linkcolor=blue,urlcolor=black}

\definecolor{cream}{RGB}{222,217,201}
\begin{document}

\pagestyle{fancy}
\thispagestyle{plain}
\fancypagestyle{plain}{
\renewcommand{\headrulewidth}{0pt}
}

\makeFNbottom
\makeatletter
\renewcommand\LARGE{\@setfontsize\LARGE{15pt}{17}}
\renewcommand\Large{\@setfontsize\Large{12pt}{14}}
\renewcommand\large{\@setfontsize\large{10pt}{12}}
\renewcommand\footnotesize{\@setfontsize\footnotesize{7pt}{10}}
\makeatother

\renewcommand{\thefootnote}{\fnsymbol{footnote}}
\renewcommand\footnoterule{\vspace*{1pt}%
\color{cream}\hrule width 3.5in height 0.4pt \color{black}\vspace*{5pt}} 
\setcounter{secnumdepth}{5}

\makeatletter 
\renewcommand\@biblabel[1]{#1}            
\renewcommand\@makefntext[1]%
{\noindent\makebox[0pt][r]{\@thefnmark\,}#1}
\makeatother 
\renewcommand{\figurename}{\small{Fig.}~}
\sectionfont{\sffamily\Large}
\subsectionfont{\normalsize}
\subsubsectionfont{\bf}
\setstretch{1.125} 
\setlength{\skip\footins}{0.8cm}
\setlength{\footnotesep}{0.25cm}
\setlength{\jot}{10pt}
\titlespacing*{\section}{0pt}{4pt}{4pt}
\titlespacing*{\subsection}{0pt}{15pt}{1pt}

\fancyfoot{}
\fancyfoot[LO,RE]{\vspace{-7.1pt}\includegraphics[height=9pt]{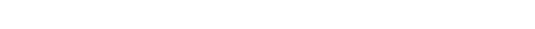}}
\fancyfoot[CO]{\vspace{-7.1pt}\hspace{13.2cm}\includegraphics{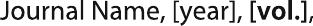}}
\fancyfoot[CE]{\vspace{-7.2pt}\hspace{-14.2cm}\includegraphics{head_foot/RF}}
\fancyfoot[RO]{\footnotesize{\sffamily{1--\pageref{LastPage} ~\textbar  \hspace{2pt}\thepage}}}
\fancyfoot[LE]{\footnotesize{\sffamily{\thepage~\textbar\hspace{3.45cm} 1--\pageref{LastPage}}}}
\fancyhead{}
\renewcommand{\headrulewidth}{0pt} 
\renewcommand{\footrulewidth}{0pt}
\setlength{\arrayrulewidth}{1pt}
\setlength{\columnsep}{6.5mm}
\setlength\bibsep{1pt}

\makeatletter 
\newlength{\figrulesep} 
\setlength{\figrulesep}{0.5\textfloatsep} 

\newcommand{\topfigrule}{\vspace*{-1pt}%
\noindent{\color{cream}\rule[-\figrulesep]{\columnwidth}{1.5pt}} }

\newcommand{\botfigrule}{\vspace*{-2pt}%
\noindent{\color{cream}\rule[\figrulesep]{\columnwidth}{1.5pt}} }

\newcommand{\dblfigrule}{\vspace*{-1pt}%
\noindent{\color{cream}\rule[-\figrulesep]{\textwidth}{1.5pt}} }

\makeatother

\twocolumn[
  \begin{@twocolumnfalse}
{\includegraphics[height=30pt]{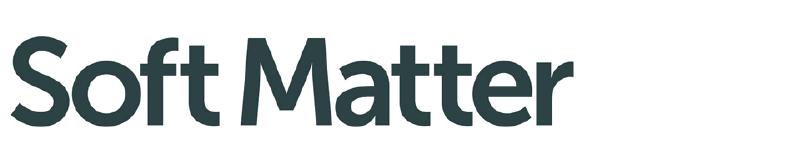}\hfill\raisebox{0pt}[0pt][0pt]{\includegraphics[height=55pt]{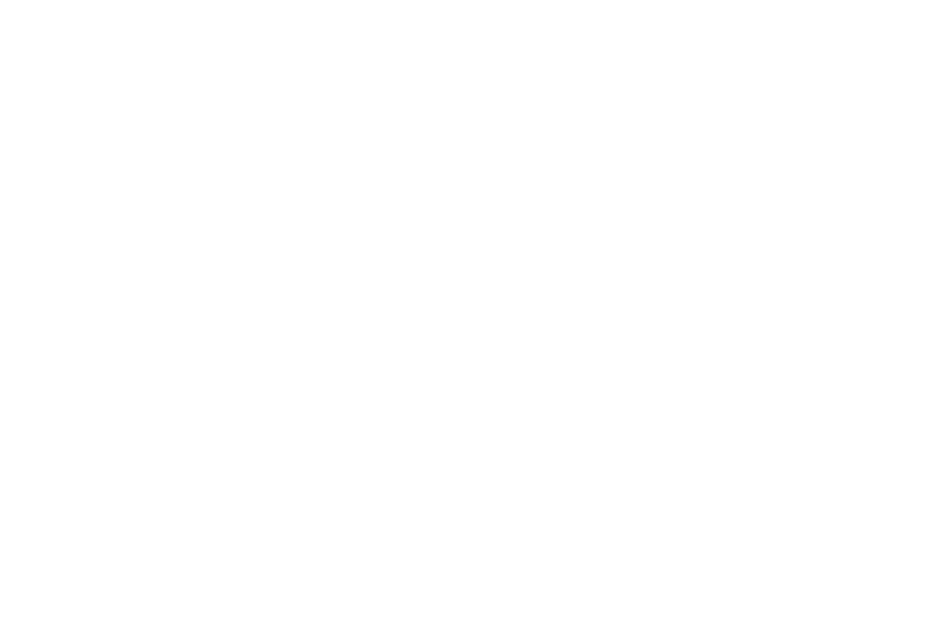}}\\[1ex]
\includegraphics[width=18.5cm]{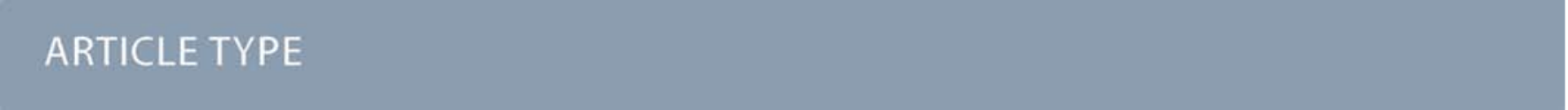}}\par
\vspace{1em}
\sffamily
\begin{tabular}{m{4.5cm} p{13.5cm} }

\includegraphics{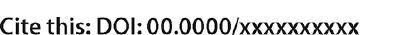} & \noindent\LARGE{\textbf{Spontaneous flows and dynamics of full-integer topological defects in polar active matter}} \\
\vspace{0.3cm} & \vspace{0.3cm} \\

 & \noindent\large{Jonas Rønning,\textit{$^{a}$} Julian Renaud,\textit{$^{b}$}   Amin Doostmohammadi,\textit{$^{\ast c}$} and Luiza Angheluta,\textit{$^{\ast \ast a}$}}\\

\includegraphics{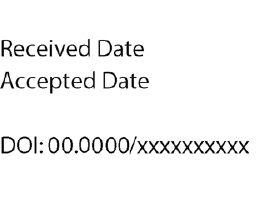} & 
\noindent\normalsize{Polar active matter of self-propelled particles sustain spontaneous flows through the full-integer topological defects. We study theoretically the effect of both polar and dipolar active forces on the flow profile around $\pm 1$ defects and their interaction in the presence of both viscosity and frictional dissipation. The vorticity induced by the active stress is non-zero at the $+1$ defect contributing to the active torque acting on the defect. A near-core flow reversal is predicted in absence of hydrodynamic screening (zero friction) as observed in numerical simulations. While $\pm 1$ defects are sources of spontaneous flows due to active stresses, they become sinks of flows induced by the polar active forces. We show analytically that the flow velocity induced by polar active forces increases away from a $\pm 1$ defect towards the uniform far-field, while its associated vorticity field decays as $1/r$ in the far-field. In the friction-dominated regime, we demonstrate that the flow induced by polar active forces enhances defect pair annihilation, and depends only on the orientation between a pair of oppositely charged defects relative to the orientation of the background polarization field. Interestingly, we find that this annihilation dynamics through mutual defect-defect interactions is distance independent, in contradiction with the effect of dipolar active forces which decay inversely proportional to the defect separation distance. As such, our analyses reveals a new, truly long-ranged mechanism for the pairwise interaction of oppositely-charged topological defects in polar active matter.} \\

\end{tabular}

 \end{@twocolumnfalse} \vspace{0.6cm}

  ]

\renewcommand*\rmdefault{bch}\normalfont\upshape
\rmfamily
\section*{}
\vspace{-1cm}


\footnotetext{\textit{$^{a}$~Njord Centre, Department of Physics, University of Oslo, P. O. Box 1048, 0316 Oslo, Norway}}
\footnotetext{\textit{$^{b}$~École Normale Supérieure, PSL Research University, 45 rue d’Ulm, 75005 Paris, France}}
\footnotetext{\textit{$^{c}$~Niels Bohr Institute, University of Copenhagen, Blegdamsvej 17, Copenhagen, Denmark}}

\footnotetext{$\dag$~Electronic Supplementary Information (ESI) available: [https://github.com/jonasron/SM\textunderscore Flowfield\textunderscore Polar]. See DOI: 10.1039/cXsm00000x/}

\footnotetext{$\ast$~Email: luizaa@fys.uio.no}
\footnotetext{$\ast\ast$~Email: doostmohammadi@nbi.ku.dk}



\section{Introduction}
Active matter refers to non-equilibrium systems of interacting, self-propelled entities that consume energy from their surrounding in the form of persistent motion and their collective interactions lead to emergent, dynamical patterns, and self-sustained flows~\cite{doostmohammadi2018active,marchetti2013hydrodynamics}. Models of active matter are largely inspired by biological systems from bacterial suspensions~\cite{doostmohammadi2018active,marchetti2013hydrodynamics} and cell monolayers~ \cite{blanch2018turbulent,lin2021energetics} and down to subcellular active systems such as mixtures of cytoskeletal filaments and motor proteins~ \cite{sanchez2012spontaneous,guillamat2017taming,Needleman17,kumar2018tunable}. However, this also pertains to non-living active systems such as layers of vibrated granular matter, microrobots, or synthetic catalytic nanomotors~\cite{kudrolli2008swarming,marchetti2013hydrodynamics}.

Several hydrodynamic models have been proposed to capture the macroscopic dynamics and emergent phenomena corresponding to a collection of active (self-propelled) particles with different symmetries and alignment interactions~\cite{kruse2005generic,marchetti2013hydrodynamics,toner1995long,juelicher2007active}. The prototypical models are based on the analogy to liquid crystals formed by rod-like particles with polar or nematic symmetries in their alignment interactions.  Active rods with only orientational alignment act as headless ``shakers" and form nematic phases described by a slowly-varying director field $\vec n$ which has head-tail (nematic) symmetry~\cite{Giomi13,marchetti2013hydrodynamics,doostmohammadi2018active,ramaswamy2003active}. By contrast, active rods that align their direction of motion tend to flock into polar systems that are described instead by a slowly-varying  polar vector field $\vec p$ ~\cite{toner1995long,kruse2004asters,kruse2005generic,juelicher2007active}. Active rod-like particles generate persistent flows sustained by the active stress originating from extensile/contractile dipolar forces. On hydrodynamic scales, this active stress is proportional to the nematic $\mathbf{Q}$ tensor order parameter, i.e. $\sigma^a =\alpha_0 \mathbf{Q}$ with a proportionality coefficient $\alpha_0$ as an effective activity parameter~\cite{marchetti2013hydrodynamics,simha2002hydrodynamic}. 
Topological defects, innate to ordered systems with broken continuous symmetries, are also present in active systems. The interplay between active stresses and nematic distortions feeds into self-sustain flows and the proliferation of topological defects to generate chaotic flows also known as active turbulence~\cite{thampi2014vorticity,doostmohammadi2017onset,chandragiri2020flow}.

From the rotational symmetry modulo $\pi$ of the $\mathbf{Q}$ tensor, it follows that the lowest energy orientational defects have a $\pm 1/2$ topological charge corresponding to $\pm \pi$ jump in the orientational phase around them. The $+1/2$ defects acquire a self-propulsion due to the net active flow passing through their cores~\cite{giomi2013defect,ronning2022flow}.  
The stable, low-energy defects in polar active systems have instead $\pm 1$ topological charges corresponding to a $2\pi$ phase jumps around the defects. This is analogous to vortices in the XY-model of 2D ferromagnets~\cite{kruse2004asters,chaikin_lubensky_1995}. 

The bulk of recent studies have focused on the formation and characterization of half-integer nematic defects~(see recent reviews~\cite{doostmohammadi2021physics,shankar2022topological}). This is in part due to the ubiquitous emergence of the nematic defects in a wide range of biological systems from subcellular filaments~\cite{decamp2015orientational,kumar2018tunable} to bacterial colonies~\cite{duclos2018spontaneous,meacock2021bacteria} and assemblies of eukaryotic cells~\cite{saw2017topological,duclos2017topological}. This is despite the fact that several biological active entities, such as bacteria or eukaryotic cells, are endowed with a clear head-tail asymmetry and directional self-propulsion, which characterizes a polar order for these systems. Such a polar order at the scale of collective is apparent from flocking domains within bacterial colonies~\cite{meacock2021bacteria} and eukaryotic cells~\cite{malinverno2017endocytic}. Nevertheless, because of the appearance of half-integer defects, at the coarse-grained level, these systems are often modelled as active nematics neglecting the polarity of the self-propelled particles. There have been proposed models that couple the evolution of polar and nematic order parameters~\cite{baskaran2008hydrodynamics,baskaran2010nonequilibrium,baskaran2012self} to allow for the coexistence of both types of symmetries.
Recently, in Ref.~\cite{Amiri_2022}, a hydrodynamic model was proposed for the coexistence of both nematic and polar alignment interactions in the polarization field $\vec p$, and was used to study different active turbulence regimes sustained by both half- and full- integer topological defects. In addition to dipolar (nematic) active forces, the model also includes a polar active force $\vec f = \alpha_p \vec p$, which describes the self-propulsion direction and is shown to suppress defect-laden active turbulence~\cite{andersen2022symmetry}. 

More recently, theoretical and experimental studies have revealed the importance of full-integer topological defects in cell assemblies. In particular, it is shown that positive full-integer defects formed due to collision of two nematic half-integer defects in fast-moving bacterial colonies can lead to the verticalization of bacteria and escape to the out-of-plane directions~\cite{meacock2021bacteria,ardavseva2022topological}. Furthermore, positive full-integer defects induced by confinement of myoblast cells in circular geometries are shown to activate cell differentiation and formation of 3D helical structures~\cite{guillamat2022integer}. A corresponding theoretical analyses, in the limit of compressible flows inside the core region of defects in small confinements, have revealed the corresponding flow fields and force patterns, and shows how confinement-induced topological defects can be used to probe the material properties of the cell layers~\cite{blanch2021integer}.

In this paper, we carry out a theoretical analysis that reveals subtle cross-talks between polar and dipolar active forces in generating spontaneous incompressible flows around both $\pm 1$ defects. We theoretically predict spontaneous vortical flows induced by dipolar active forces around $+1$ defects which correspond to isotropic active stress and pressure fields. Interestingly, the competition between dipolar force and viscous force leads to a flow reversal close to the core of the $+1$ defect. This effect is also confirmed by numerical simulations of the full hydrodynamic model including additional passive stresses. However, it turns out the hydrodynamic screening induced by friction lifts up this flow reversal effect. By contrast, the $-1$ defects which have an innate $4$-fold symmetry in the polarity field lead to $8$-fold symmetries of 
the active flows induced by dipolar active forces. The same $8$-fold symmetry is present also in the profile of the pressure field. We demonstrate that polar active force trigger a distinct active flow pattern characterised by uniform flow in the far-field of its source, i.e. $\pm 1$ defects. This is very different that the active flows sustained by dipolar forces which decay algebraically with distance. Polar active forces have a drastic effect on the mutual interactions between defects by promoting fast annihilation of defect pairs through non-local and non-reciprocal attraction forces. We show that this distance-independent mechanism of defect pair annihilation is responsible for the suppression of active turbulence by polar forces as recently reported in Ref.~\cite{andersen2022symmetry}.  

The paper is organized as follows: We begin in Section~\ref{sec2} by introducing the flow equations within a minimal polar hydrodynamic model and derive the corresponding defect kinematic equations using Halperin-Mazenko formalism~\cite{angheluta2021role}. The main analytical results on the active flow velocity induced around isolated $\pm 1$ are discussed in Section~\ref{sec3}. We also compare the analytical predictions with direct numerical simulations and find very good agreement of the flow profiles around defects and the flow reversal pattern near the $+1$ defect cores. 
In Section~\ref{sec4}, we discuss the polar active force and its effect on the defect kinematics.
In particular, we consider the motion of a pair of oppositely charged full-integer defects under the polar active flows induced by each defect and demonstrate that polar active forces enhance the defect annihilation rate. In a recent study Ref.~\cite{andersen2022symmetry}, it was numerically evidenced that polar active forces suppress defect-laden active turbulence and tend to restore polar order. Here, we demonstrate theoretically that polar active forces have a net effect on the defect kinematics to promote defect pair binding and subsequent annihilation of defects of opposite topological charges. The final section provides a summary of the theoretical insights and concluding remarks.

\section{Hydrodynamic model of polar active matter} \label{sec2}

We consider the hydrodynamic model~\cite{Amiri_2022} that describes the collective dynamics of self-propelled entities in terms of the evolution of the polar order parameter $\vec p$ given by %
\begin{align}
     \partial_t \vec p + \vec u \cdot \nabla \vec p + \lambda \mathbf E \cdot \vec p     + \mathbf \Omega \cdot \vec p &= -\frac{1}{\gamma} \frac{\delta\mathcal F}{\delta \vec p},
     \label{eq:p}
\end{align}
where $\gamma$ is the rotational viscosity, $\lambda$ is the flow-aligning parameter~\cite{chandragiri2019active}, and the free energy favoring the polar order is described as
\begin{align}
     \mathcal F=\int d\vec{r} \left\{A\left(-\frac{|\vec p|^2}{2}+\frac{|\vec p|^4}{4}\right)
     +\frac{K_p}{2}|\nabla\vec{p}|^2\right\}.\label{eq:F}
\end{align}

Here, $K_p$ is the isotropic elastic constant for distortions in the polarity field and $A$ is the height of local energy barrier. The polarity is coupled with the flow field $\mathbf u$ which is described by the incompressible Stokes equations 
\begin{align}
    (\Gamma-\eta \nabla^2)  \vec u  &=   \alpha_0 \nabla\cdot \mathbf{Q}-\nabla  P, 
    \label{eq:Stokes} \\
    \nabla \cdot \vec u &= 0,
     \label{eq:incompressibility}
\end{align}
where the dipolar active forces are mediated by the nematic order through the nematic tensor $Q_{ij} = (p_i p_j - \frac{p^2}{2} \delta_{ij})$ and the activity parameter $\alpha_0$. The incompressibility constraint determines the fluid pressure  
$P$. The viscosity is set by $\eta$ while the friction with a substrate is introduced by the frictional drag $\Gamma$~\cite{doostmohammadi2016stabilization}. Since we focus on theoretical derivations of the \textit{active flows} induced by polar and dipolar forces, we hereby neglect the additional passive stresses which depend on the polarity and that are typically present in the Stokes equations simulated numerically. 

We consider the dimensionless forms of these equations following the appropriate rescalings in units of length $\xi = \sqrt{K_p/A}$ and time $\tau = \gamma/A$. The system is then controlled by three dimensionless parameters: the scale number $\zeta = \ell_d/\xi$  is the ratio between the hydrodynamic dissipation length $\ell_d = \sqrt{\eta/\Gamma}$ and the nematic coherence length $\xi$, and the rescaled activity $\tilde{\alpha}_0=\alpha_0 \tau /(\xi^2) =\alpha_0 \gamma/(K_p)$. Thus, the dimensionless flow equations read as
\begin{align}\label{eq:flow_eqs}
    (1-\zeta^2 \nabla^2)  \vec u  = \vec F_a-\nabla  P ,
    \\
    \nabla^2 P = \nabla \cdot \vec F_a,
\end{align}
where the dipolar active force induced by the active stress is
\begin{equation}
    \vec F_a = \frac{\tilde{\alpha}_0}{\Gamma} \nabla\cdot \mathbf Q.
\end{equation}
For infinite systems, the flow velocity and pressure originated from dipolar active forces can be calculated from convolution integrals of the source terms and the corresponding Green's functions as
\begin{align}
    P(\vec r) &=\frac{1}{2\pi} \int d\vec{r'} \ln(|\vec{r}-\vec{r'}|) \left[\nabla' \cdot \vec F_a\right],
    \label{eq:Greens_Pressure}
\end{align}
and
\begin{align}
    \vec u(\vec r) &=\int \frac{ d\vec{r'}}{2\pi\zeta^2} K_0\left(\frac{|\vec r - \vec{r'}|}{\zeta}\right) \left(\vec F_a(\vec{r'})- \nabla'P(\vec{r'}) \right).
    \label{eq:Greens_Flow}
\end{align}

\begin{figure*}[ht]
    \centering
    \includegraphics[width = 0.45\textwidth]{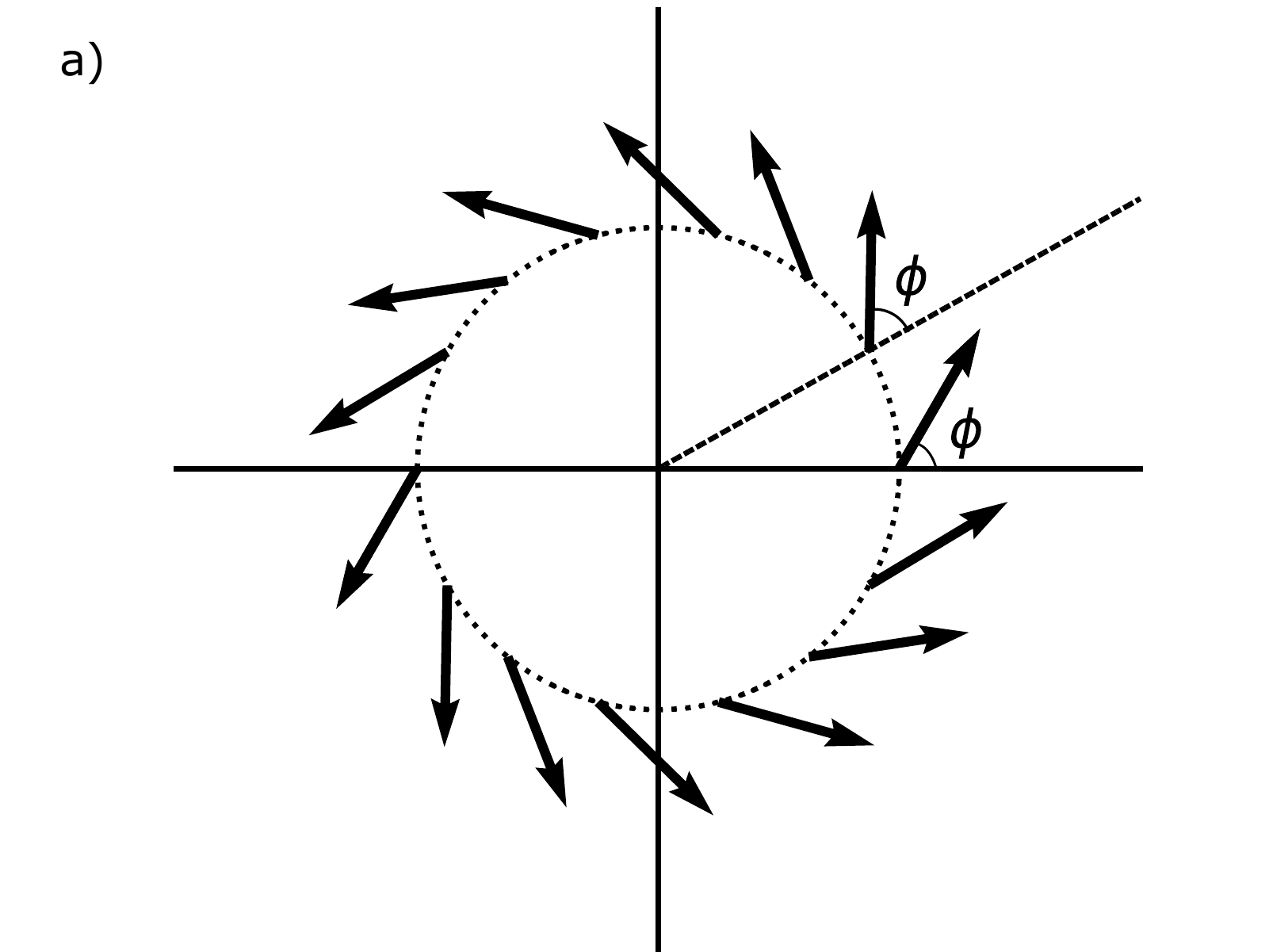}
    \includegraphics[width = 0.45\textwidth]{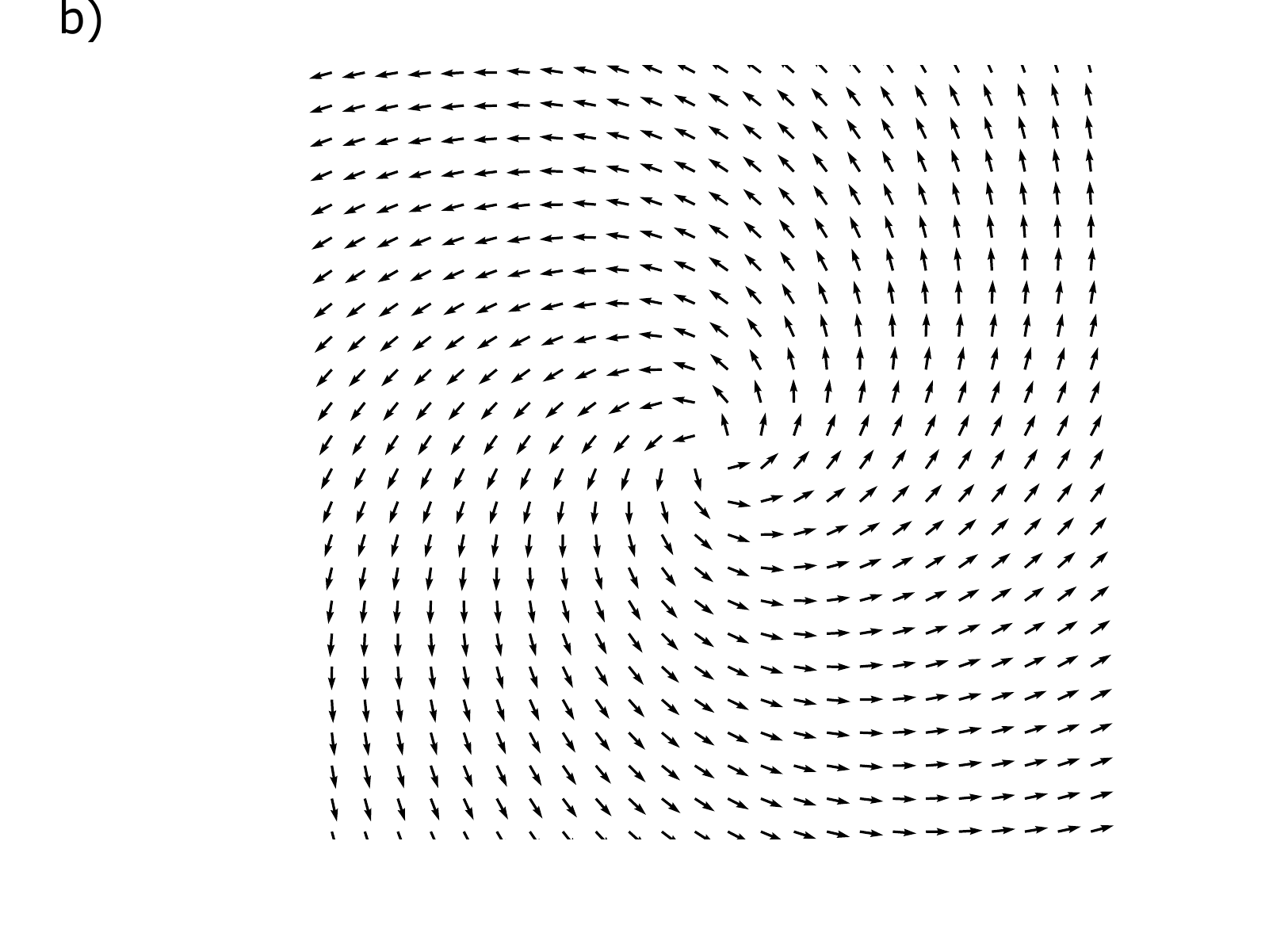}
    \includegraphics[width = 0.45\textwidth]{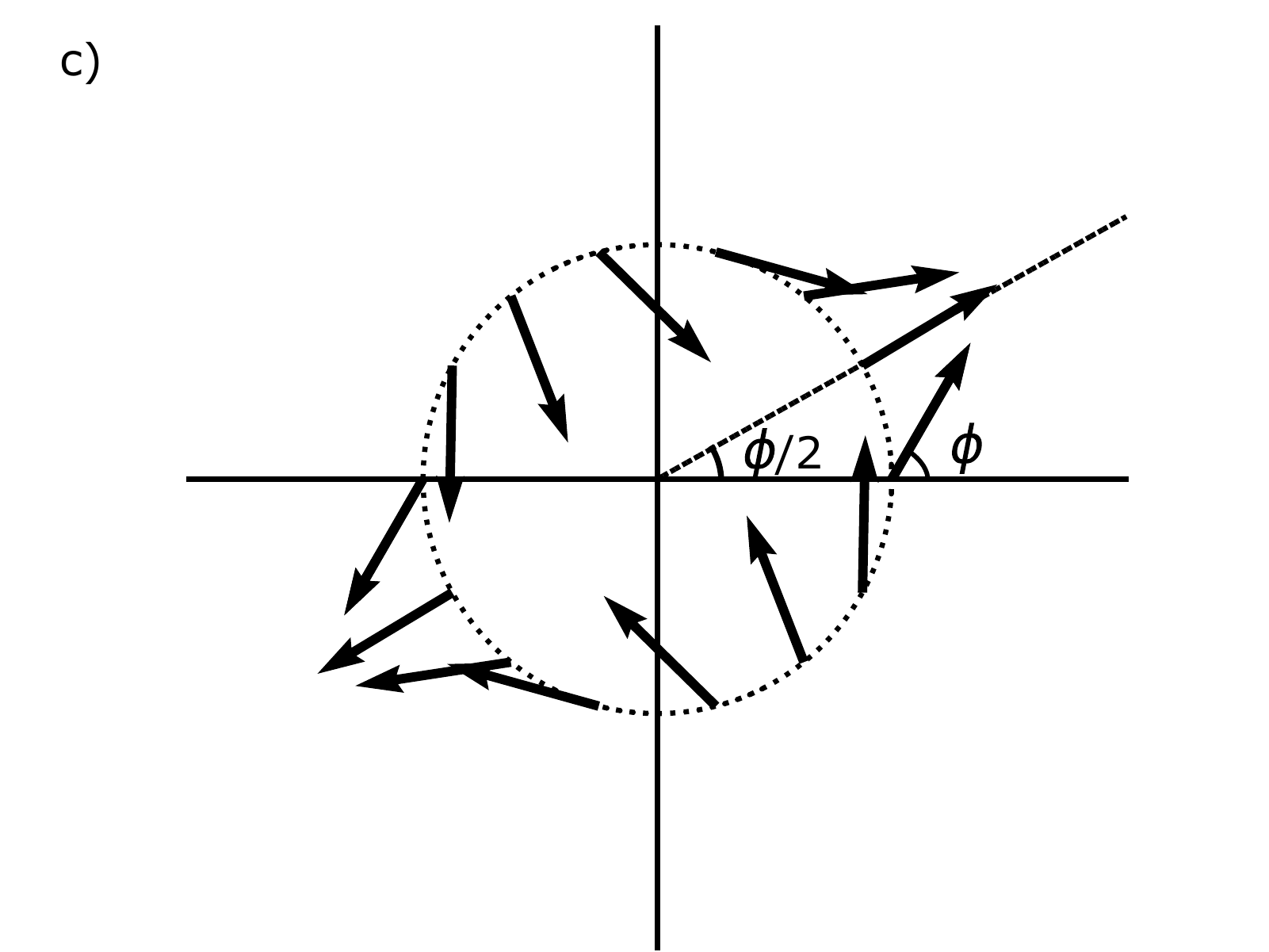}
    \includegraphics[width = 0.45\textwidth]{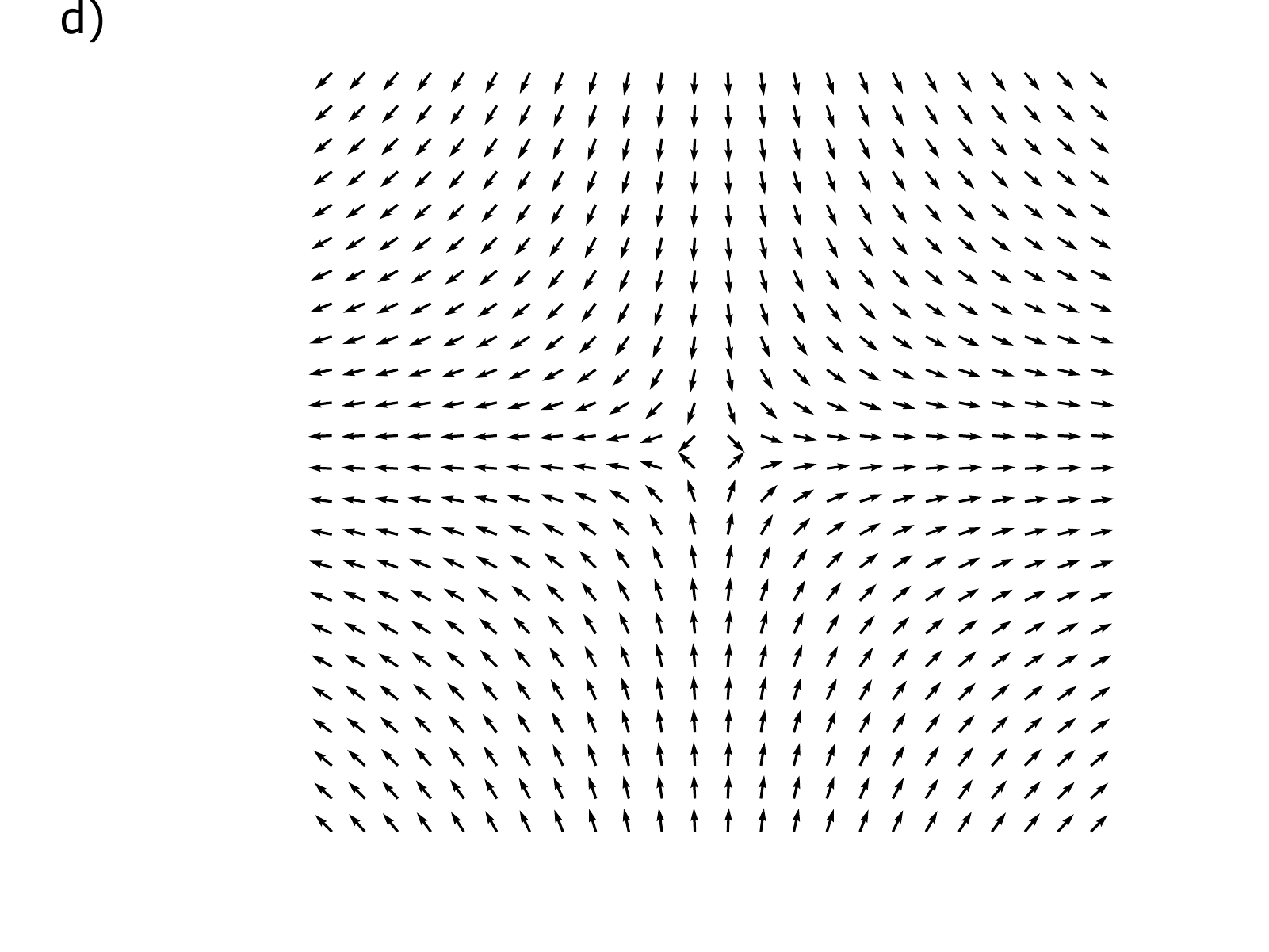}
    \caption{\textbf{Polarity field around the $\pm 1$ topological defects}. a) and b) show a $+1$ defect, while c) and d) illustrate a $-1$ defect. The drawings a) and c) show the $\vec p$ plotted on a circle with the angle $\phi$ marked. This illustrates why $\phi$ can be set to zero by a change of basis for the $-1$ defect and not for the $+1$ defect. b) and d) show the polarity fields around a $+1$ defect with $\phi=\pi/3$ (spiral) and a $-1$ defect with $\phi=0$ respectively (Changing $\phi$ rotates the negative defect).}
    \label{fig:defect_streamlines}
\end{figure*}

To derive analytical expressions, we evaluate the source term due to the active stress using the order parameter for an isolated point defect located at the origin and given by 
\[\vec p(r,\theta) = \chi(r)[\cos(q\theta +\phi) \vec e_x+ \sin(q\theta +\phi)\vec e_y],\]
where $\theta = \arctan(y/x)$ is the polar angle in a coordinate system centered at the defect position and $\chi(r)$ is the core function, where we assume that the core size is much smaller than any other length scales and set $\chi(r)=1$. 
The angle of the polar vector order parameter for an ideal defect is given by $ \Theta(\theta) =q\theta + \phi$ \cite{chaikin_lubensky_1995}, where $q\theta$ is the singular part giving the winding number $q$ when performing the integral $\oint d\Theta=2\pi q$ on a contour surrounding the defect and $\phi$ is a constant.
For defects of charge $q \neq +1$ this constant sets the orientation of the defect and can be ignored since we can always transform to a system with $\phi=0$ by a change of basis.
When $q =+1$ it is impossible to remove a non-zero $\phi$ by changing the basis~\cite{tang2017orientation}. 
The positive defects are rotationally symmetric and the constant $\phi$ becomes important since it distinguishes different types of defects: $\phi =0$ corresponds to an aster and $\phi = \pi/2$ gives rise to a vortex, and any value in-between corresponds to a spiral defect. For a more intuitive depiction of this, we plot in Fig.~\ref{fig:defect_streamlines} a) and c) the $\vec p$ field on the circumference of a circle centered at $\pm 1$ defects to show that $\phi$ corresponds to the constant angle that $\vec p$ makes with the radial direction. For the negative defect one can define a polarisation from the line where $\vec p$ is pointing radially outward from the core, the angle of this polarisation is half of $\phi$. 
The different patterns of the $\vec p$ field around a $\pm 1$ defect are also illustrated in Fig.~\ref{fig:defect_streamlines} b) and d).

We use the Halperin-Mazenko formalism~\cite{mazenko1997vortex,halperin1981published} for tracking topological defects as zeros of the polar order parameter to derive the corresponding equations of motion of defects from the evolution of the $\vec p$ field similar the approach from Ref.~\cite{angheluta2021role} to describe orientational defects in active nematic films. The basic idea is that since the $\pm 1$ defects are associated simultaneously with topological singularities in the orientation field and zero magnitude of the vector order parameter, we can track their position by Dirac delta functions centered at the zeros of the $\vec p$ field. Hence, a configuration of well-separated defects punctuating the $\vec p$-field corresponds to a defect charge density field, which can be written equivalently either as a superposition of the delta functions associated with the topological singularities located in the physical space or as the zeros in the order parameter space, 
\begin{equation}
    \rho(\vec r, t) = \sum_n q^{(n)}\delta(\vec r -\vec r^{(n)}) = D\delta(\vec p),
\end{equation}
where $D$ is the determinant of the polarity distortion tensor $\nabla \vec p$, i.e. $D = \epsilon_{ij}\partial_i p_x \partial_j p_y$ which can be expressed equivalently in the complex representation $\psi = p_x +ip_y$ as $2i D = \epsilon_{ij}\partial_i \bar\psi \partial_j \psi $. The $D$-field is zero in regions of uniform polar order and becomes nonzero where there are distortions in the orientation field.  For configurations of well-separated defects punctuating a uniform polar order state, the $D$ field is zero everywhere except at the defect positions labeled by the index $n$, where the topological charge $q_n = \pm 1$ is determined by the sign of $D$. Thus, the $D$ field represents a non-singular charge density field. 

It can be shown that the $D$-field follows the conservation law~\cite{mazenko1997vortex}
\begin{equation}
    \partial_t D +\partial_i \vec J_i =0,
\end{equation}
with the current density $J = \epsilon_{ij}\Im(\partial_t \psi \partial_j \bar\psi)$ determined by the evolution of the polar order. Combining this with the conservation of topological charge density $\rho$, we can derive a general expression for the defect velocity in terms of the $D$ and its current, $\vec v_n = \vec J/D |_{\vec r_n}$~\cite{angheluta2021role}.  We parameterise the polarity field from the viewpoint of the $n$-th defect as $\psi = \psi_n e^{i\phi_n}$, where $\phi_n$ is the smooth phase at the defect position $\mathbf r_n$ and $\psi_n = |\vec r-\vec r_n| e^{i\theta_n}$ using that the polarization vanishes linearly while its phase $\theta_n$ is singular at the pointwise defect position~\cite{mazenko2001defect,pismen1999vortices}.  
Within this approach, the general expression of the defect current density can be reduced to a closed expression for the defect velocity given by  
\begin{align} \label{eq:defect_velocity}
    \vec v_n = \vec{u}(\vec r_n) + 2q_n\nabla^\perp \phi_n\bigg|_{\vec r= \vec r_n}.
\end{align}

Using the stationary phase approximation, i.e. that the constant (equilibrium) phase remains stationary when it is punctuated by moving defect singularities, we can further simplify the defect velocity and express it in terms of the net spontaneous flow and the forces induced by the other defects as~\cite{vafa2020multi,angheluta2021role}
\begin{equation}\label{eq:point_vortex_model}
    \vec v_n = \vec u(\vec r_n) + 2\sum_{k \neq n} q_n q_k\frac{\vec r_n -\vec r_k}{|\vec r_n -\vec r_k|^2}.
\end{equation}
The topological defects interact through Coulomb-like forces, where like-signed defects repel, and opposite-signed defects attract each other. However, there are additional interactions through the flow field $\vec u$ which depends on the dipolar active forces. In the next section, we derive analytic expressions for this flow velocity and discuss its effect on the defect motion.

\section{Active flow fields around $\pm 1$ topological defects}\label{sec3}
From the parameterization of the $\vec p$-field for a pointwise defect, we can evaluate the dipolar active force induced by an isolated defect with charge $q= \pm 1$ in an otherwise homogeneous polarity field with constant background orientation $\phi$ as
\begin{align}
\vec F_a^{+} = \frac{\tilde{\alpha}_0}{\Gamma r^2} \left[\vec r \cos(2\phi) - \vec r^\perp \sin(2\phi)\right], \\
 \vec F_a^{-} = -\frac{\tilde{\alpha}_0}{\Gamma r}\mathbf{R}_{2\phi} \left[\cos(3\theta)\vec e_x -\sin(3\theta) \vec e_y\right],
 \end{align}
where $\vec r^\perp = (y,-x)$, and $\mathbf{R}_{2\phi}$ is a matrix that rotates the vector by $2\phi$, and can be removed by a change of basis. 
Notice that the first term in $\vec F_a^+$ is a source of gradient flow, which however is removed by pressure through the incompressibility condition. The second term proportional to $\sin(2\phi)$ is related to a rotated gradient and induces a purely \textit{vortical} flow.

\begin{figure*}[ht]
     \centering
     \includegraphics[width = 0.4\textwidth]{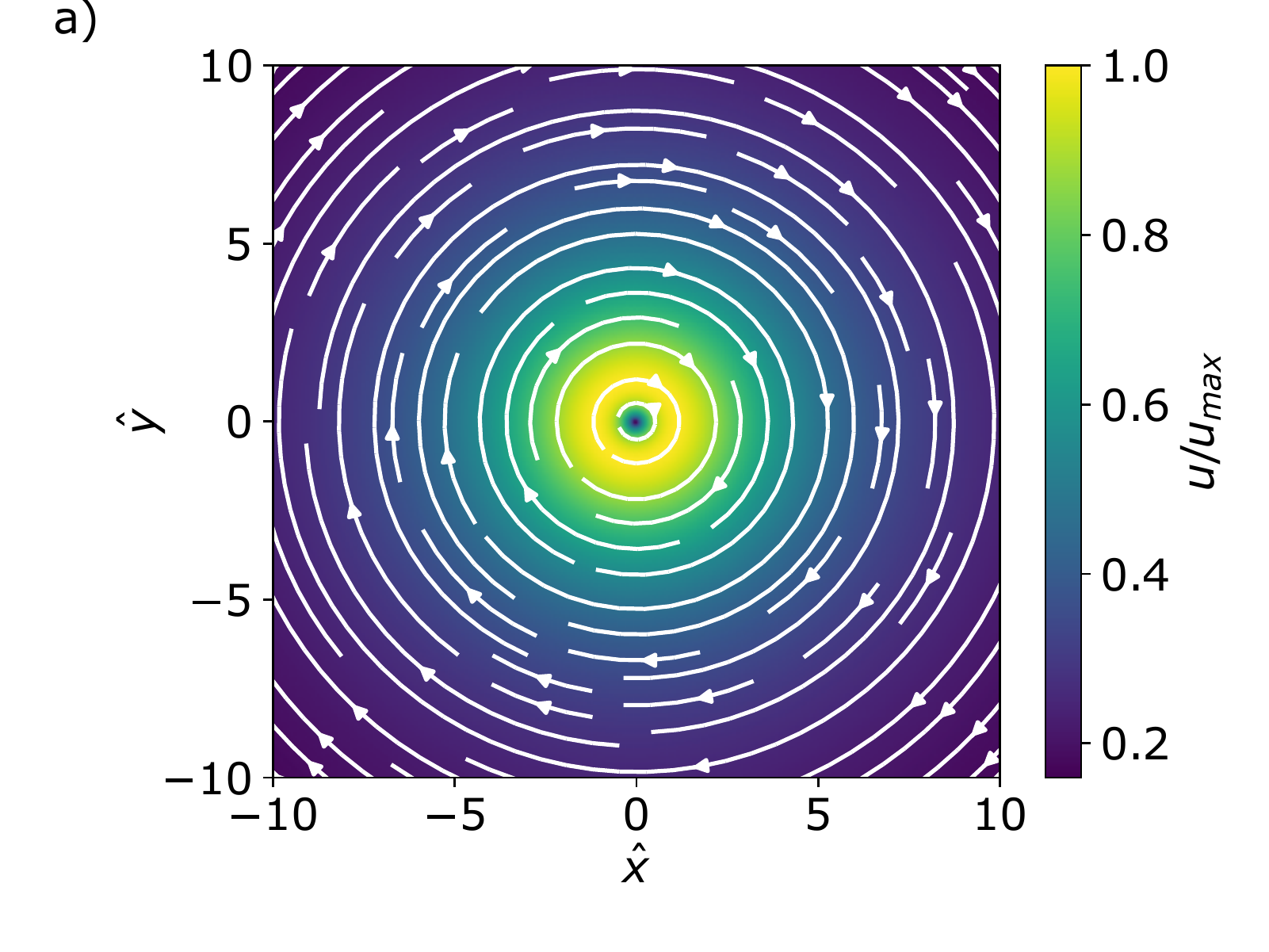}
     \includegraphics[width = 0.4\textwidth]{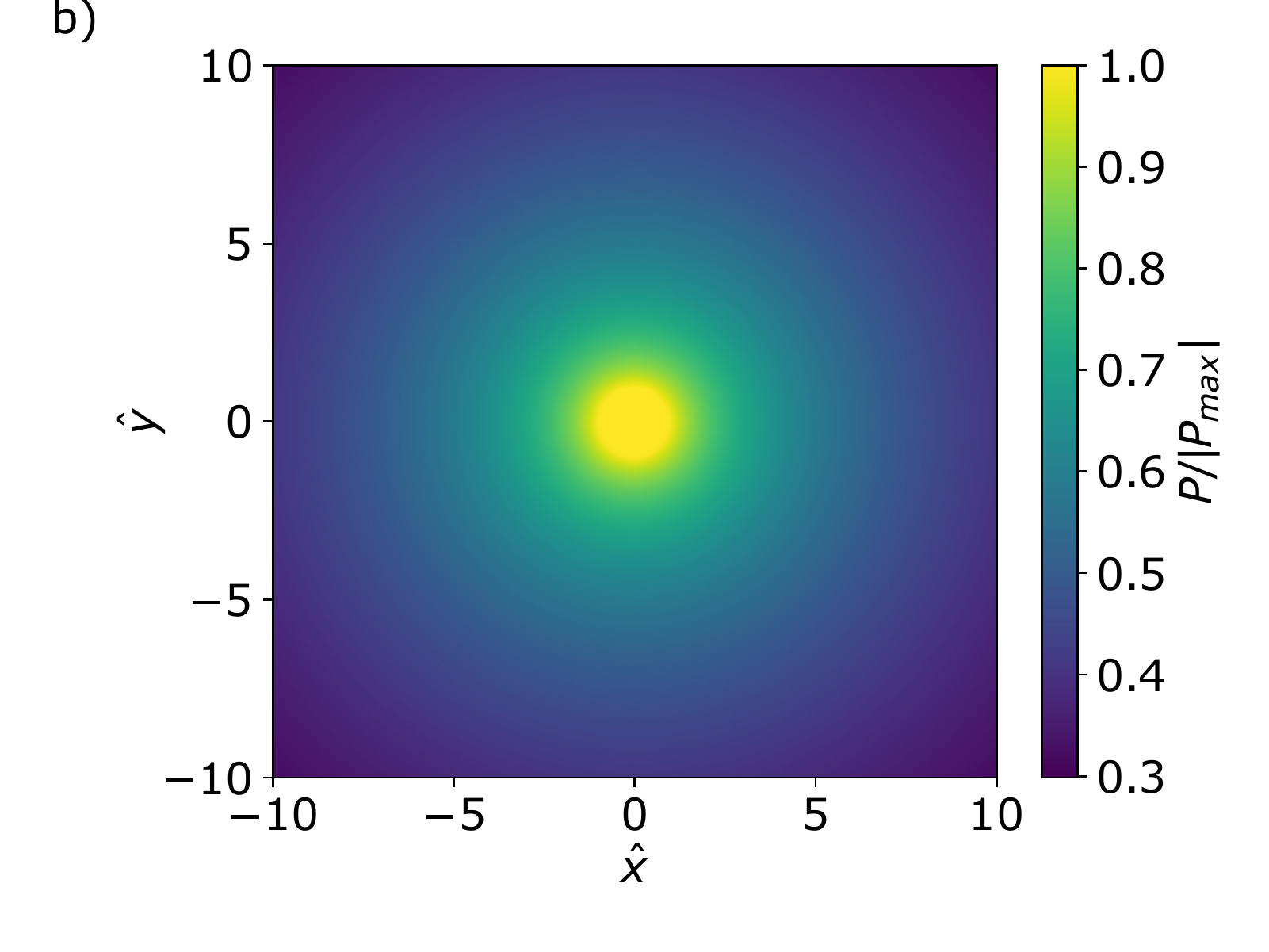}
     \includegraphics[width = 0.4\textwidth]{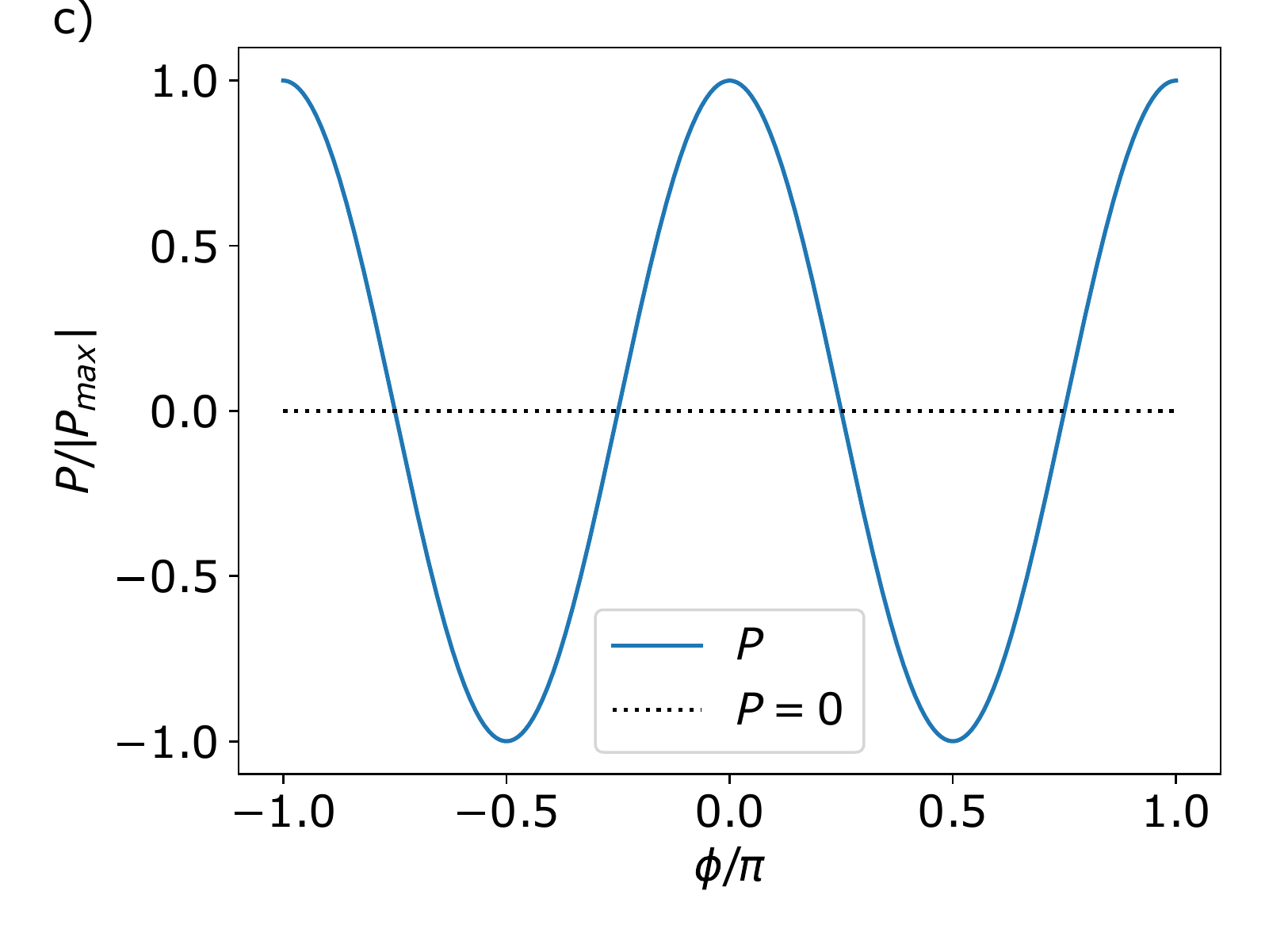}
     \caption{Incompressible active flow streamlines (a) and corresponding pressure field (b) around an isolated $+1$ vortex with $\sin(2\phi) > 0$ in an extensile ($\tilde{\alpha}_0 <0$) system. The colormap shows the magnitude of the velocity and pressure normalized by the maximum. Note that the pressure in the far-field diverges with the system size $L$. Since the pressure diverges at the centre the pressure inside the core $r<1$ has been set to be constant. c) shows the normalised pressure, for an extensile system, at a fixed radius as a function of $\phi$. Notice that for $\phi = \pm\pi/4 +\pi n$ the pressure vanishes.
     }
     \label{fig:vortex_stream_pos}
 \end{figure*}

We insert these forces into the integrals in Eq.~(\ref{eq:Greens_Pressure}) and~(\ref{eq:Greens_Flow}), and solve them as described in the Supplementary Material~\cite{SI}. The resulting expressions for the active flow velocity and pressure for a $+1$ defect reduce to 
 \begin{align}
     u_a^{+}(r,\theta) &= \frac{\tilde{\alpha}_0}{\zeta \Gamma \hat r} (1- \hat r K_1(\hat r)) e^{i(\theta+\pi/2)} \sin(2\phi),\label{eq:pos_vel} \\
    P_a^+ &= 
    - \frac{\tilde{\alpha}_0}{\Gamma } \cos(2\phi) \ln{\frac{L}{r}}. 
    \label{eq:pos_pres}
 \end{align}
Here $L$ is a cutoff scale set by the system size. We have written the velocity field as a complex field defined as $u = u_x +i u_y$, where $r$ and $\theta$ are the polar coordinates centered at the defects position. We define the scaled radial coordinate as $\hat r = r/\zeta$, which is the same as changing the length scale to the hydrodynamic dissipation length $\ell_d$ (see Fig.~\ref{fig:vortex_stream_pos}, with the cutoff scale $L=50$). Notice that the pressure is \textit{isotropic} and its gradient force cancels the radial component of the dipolar active force, thus no net pressure flow. Furthermore, the corresponding vorticity is also \textit{isotropic}  
\begin{equation}
    \omega^+_a= \frac{\tilde{\alpha}_0}{\Gamma \zeta^2}K_0(\hat r)\sin(2\phi),
\end{equation}
and is non-zero at the centre of the defect as evidenced also in Fig.~\ref{fig:vortex_stream_pos}. This has important consequences for the stability of vortex, spiral, or aster shaped positive defects as discussed in detail in Ref.~\cite{kruse2004asters}. The sign of the global circulation is modulated by the character of the $+1$ defect through $\phi$. 

Similarly, we find analytical expressions for the active flow velocity, pressure and vorticity related to the $-1$ defect, which are written in a compact form as
 \begin{align}
    u^{-}_a(r,\theta) &= -\frac{\tilde{\alpha}_0}{2\Gamma \zeta} \left(f_3^a(\hat r) e^{-3i\theta} + f_5^a(\hat r)e^{5i\theta}  \right),\\
       P_a^- &= 
   -\frac{\tilde{\alpha}_0}{4\Gamma  r^4} (x^4-6x^2y^2 +y^4), \\
   \omega_a^-(r,\theta) &= -\frac{\tilde{\alpha}_0}{\Gamma \zeta^2} \sin{4\theta} f_\omega^a(\hat r).
 \end{align}
The functions $f_3^{a}$, $f_5^{a}$, and $f_{\omega}^{a}$ giving the radial dependence of the velocity and vorticity are listed in the Appendix.  
  \begin{figure}[ht]
     \centering
       \includegraphics[width=0.4\textwidth]{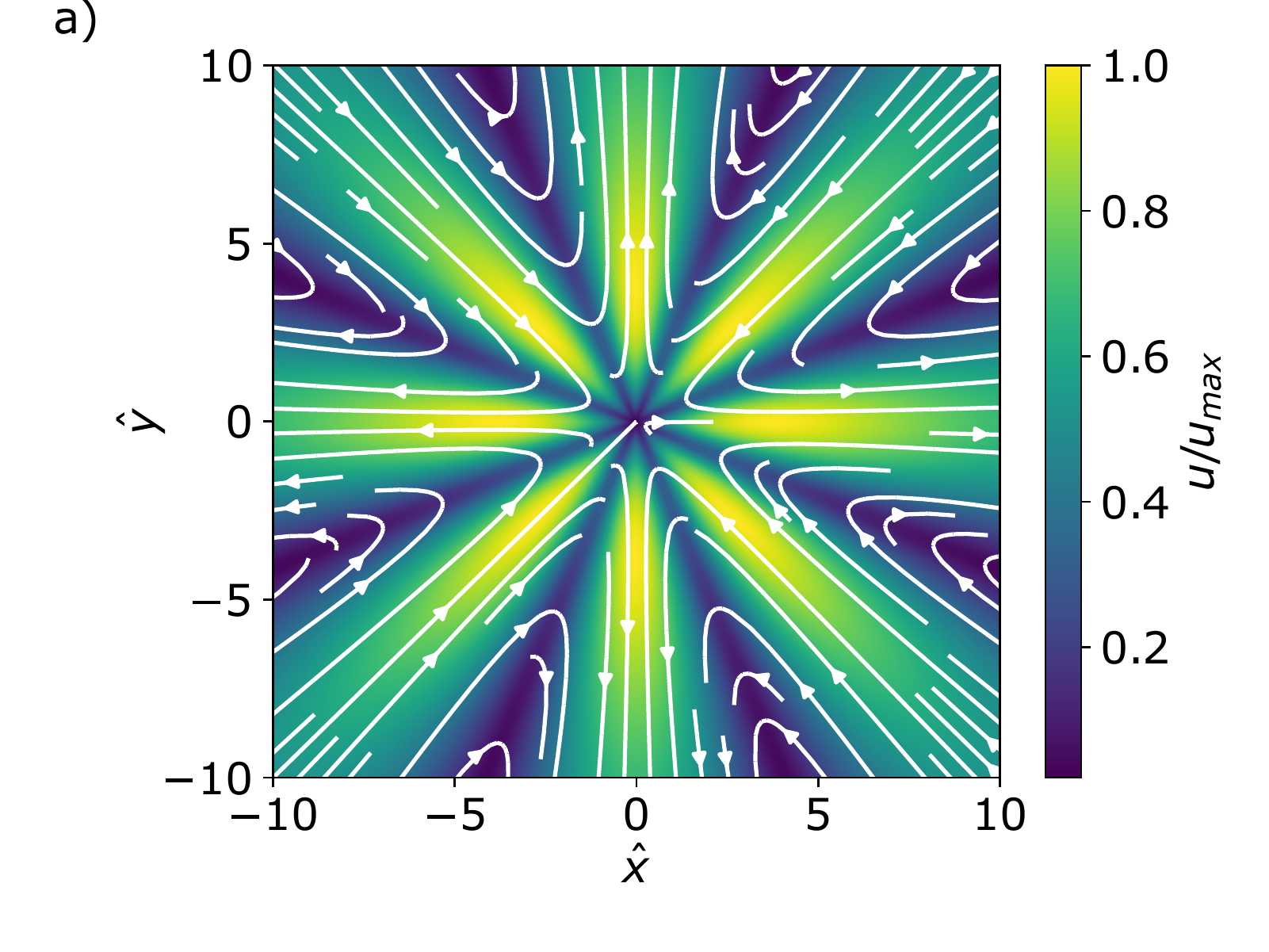}
      \includegraphics[width=0.4\textwidth]{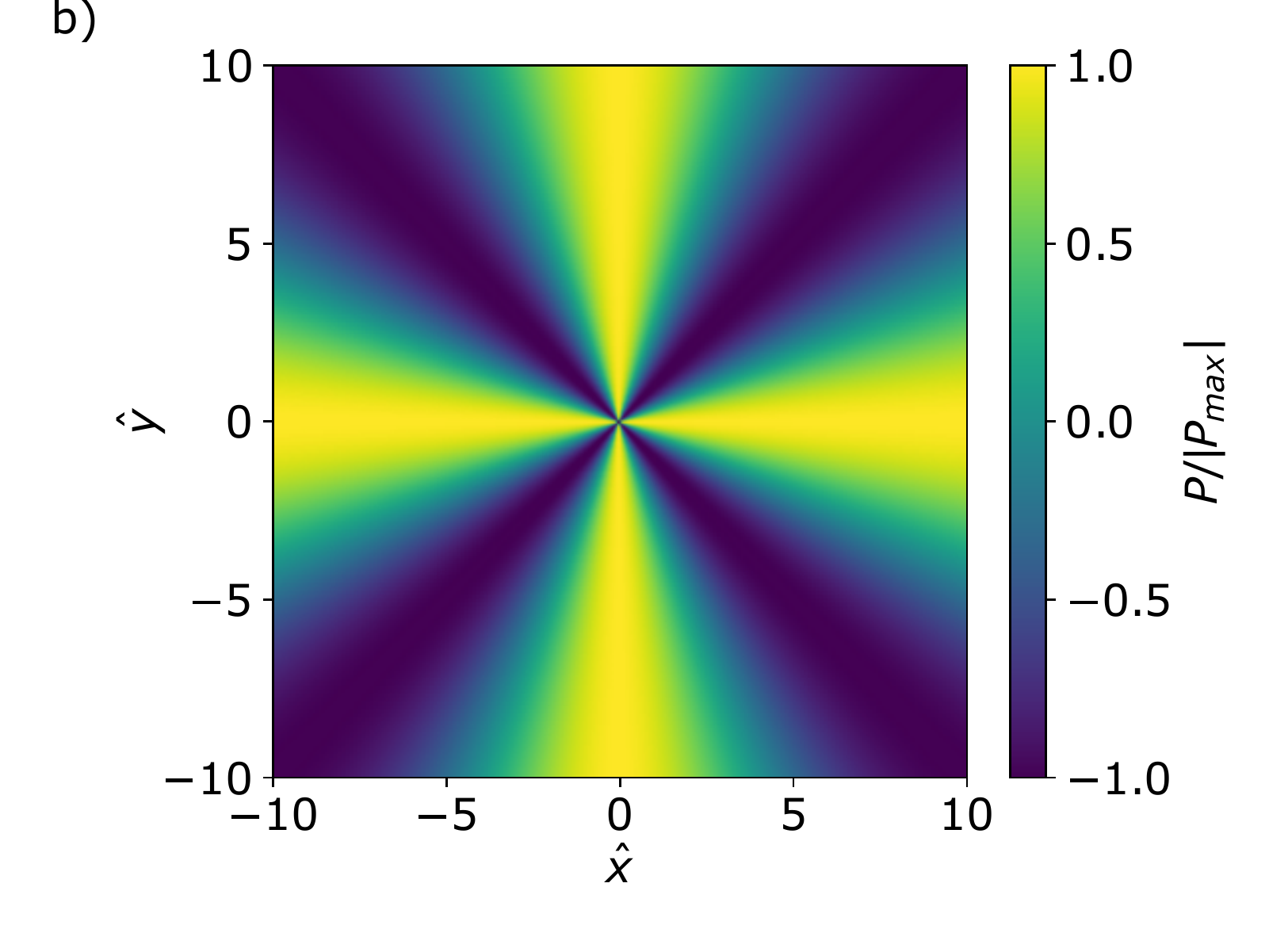}
     \caption{Incompressible active flow streamlines (a) and corresponding pressure field (b) around an isolated $-1$ vortex with $\phi = 0$ in an extensile ($\tilde{\alpha}_0 <0$) system. The colormaps show the normalised magnitude of velocity (a) and pressure (b). 
     }
    \label{fig:vortex_stream_neg}
 \end{figure}
 These fields are also plotted in Fig.~\ref{fig:vortex_stream_neg} showing that while the positive defect has closed streamlines also in an infinite system, the negative defect has an 8-fold symmetry of the vorticity, underlying a 4-fold symmetry of the polarity field around the defect. The streamlines do not close in an infinite system, but in confinement they might close due to boundary conditions or influence from other defects similarly to the vortices formed around $\pm 1/2$ defects in an active nematic \cite{giomi2014defect,ronning2022flow}. It is important to note the distinctions of the analytical descriptions provided herein with a recent calculations of {\it compressible} flow fields inside circular confinements~\cite{blanch2021integer}. In the latter, the analyses is restricted to circular domains with radius smaller than the coherence length $R^2 \ll K_p / A$, i.e the limit where relaxation/penetration length in the free energy is larger than the system size. Those results are therefore in the opposite limit of the calculation performed here.

 In addition to the characteristic flow fields, our closed form analytic descriptions provide an insight into the isotropic stress patterns, i.e., half of the trace of the stress tensor, around full-integer topological defects. Such isotropic stresses have been shown to be a determining factor for the biological functionality of nematic defects, where concentration of compressive stress around $+1/2$ defects was shown to lead to cell death and extrusion~\cite{saw2017topological}, while the tensile stresses around $-1/2$ defects has been shown to lead to spontaneous gap opening in epithelial cell layers~\cite{sonam2022mechanical}. Similarly, we find distinct isotropic stress patterns around positive and negative full-integer defects: while around a negative ($-1$) defect alternating regions of tension and compression appear in a 4-fold symmetric pattern (Fig.~\ref{fig:vortex_stream_neg}b), a strong augmentation of compressive stress is observed at the core of positive $+1$ defects (Fig.~\ref{fig:vortex_stream_pos}b). We conjecture that such a concentrated compressive stress at $+1$ defects could contribute to the activation of mechanosensitive signals in cell layer and potentially be linked to the recent observation of the cellular differentiation at +1 defect cores in cartilage cells~\cite{makhija2022topological}.
\subsection{Vortical flow reversal around $+1$ defect}
\begin{figure}[ht]
     \centering
     \includegraphics[width = 0.4\textwidth]{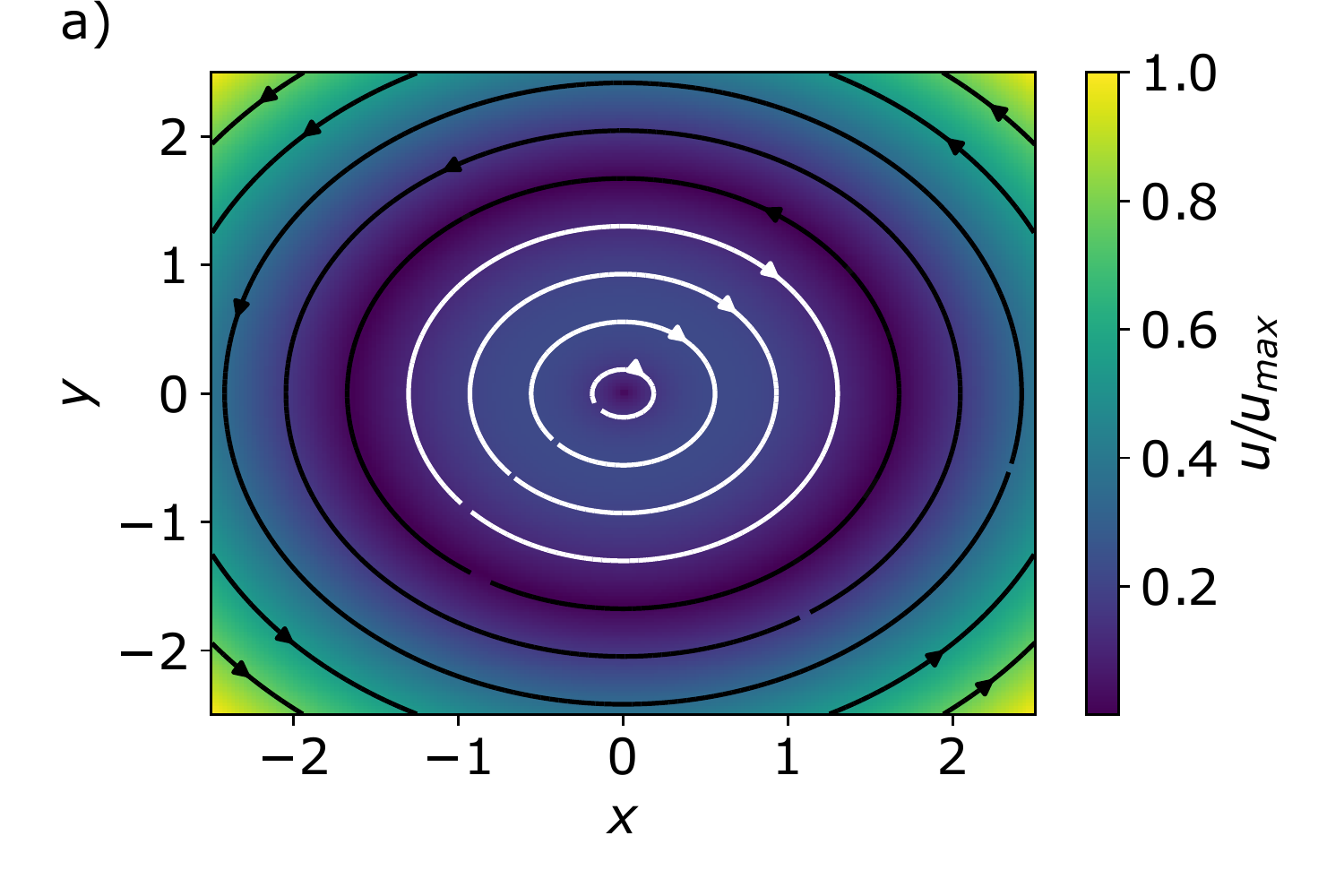}
     \includegraphics[width = 0.4\textwidth]{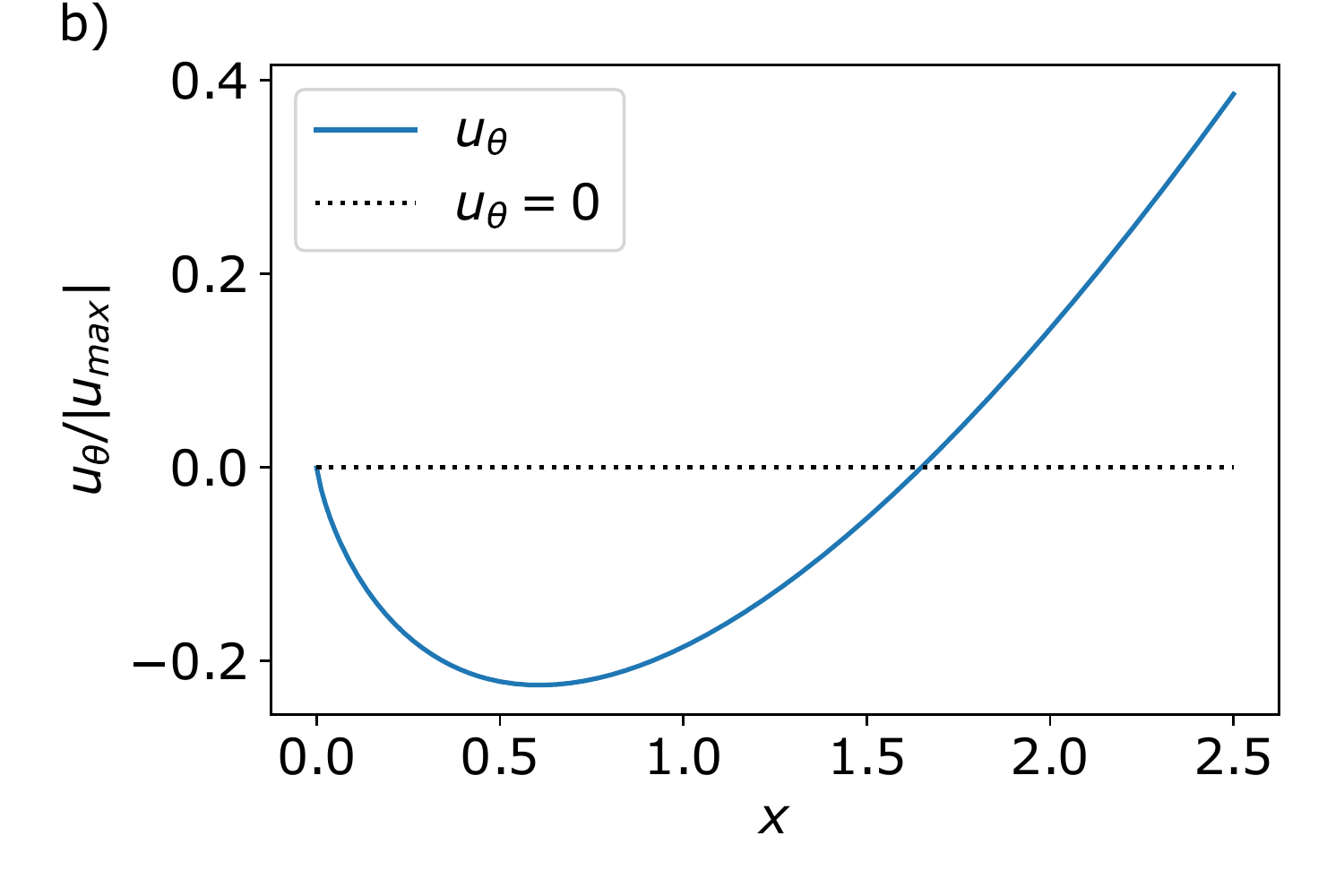}
     \caption{(a) Streamlines of the theoretical active flow around an isolated $+1$ vortex for $\sin(2\phi) > 0$ for an extensive system and $\Gamma = 0$. The colormap shows the flow magnitude. The change in color of the streamlines from white to black highlight the flow reversal at $r\approx 1.7$. This is more apparent in (b) which is a cross section of the azimuthal velocity component along the positive $x$ axis. The velocity changes sign when it crosses the zero line (doted black).    }
     \label{fig:vortex_stream_frictionless}
 \end{figure}

The results presented above provide closed-form analytical expressions for incompressible flows around full-integer topological defects in polar active matter. Interestingly, in the viscous-dominated regime ($\Gamma =0$), we find that the activity-induced flow around the $+1$ defect forms two counter-rotating vortices around the defect core, and thus the rotational flow reverses sign on a length scale comparable with the coherence length outside of the defect core. 
In this limit, the flow velocity becomes instead
\begin{align}
    u_a^+  = \frac{i\tilde{\alpha}_0}{4\tilde \eta }r(1-2 \ln(r)) e^{i\theta } \sin(2\phi),
\end{align}
and its corresponding vorticity is thus 
\begin{equation}
    \omega^+_a = -\frac{\tilde{\alpha}_0}{\tilde\eta} \sin(2\phi)\ln(r),
\end{equation}
where $\tilde \eta = \eta/\xi^2$.
We notice that the vorticity changes handedness for $r = \sqrt{e}  \approx 1.7$ giving rise to two counter-rotating regions as shown in Fig.~\ref{fig:vortex_stream_frictionless}. Notice that the flow diverges at $r\rightarrow \infty$ since there is no screening lengthscale. The mathematical reason for this flow reversal is that the logarithmic Greens function changes sign when $r<1$,  unlike the Bessel function $K_0$ kernel which appears when both friction and viscosity are present. Note that in the absence of friction, the flow equations have no natural lengthscale, thus the rescaling must be done with an appropriate lengthscale coming from the free energy, such as the coherence length which is consistent with the numerical observations of the flow reversal discussed next. Similar flow reversal is also expected in systems confined to circular domains that are small enough such that the energy scale can be treated as linearly dependent on the radial distance from the center of singularity $\chi(r)\sim r$ in the entire domain~\cite{blanch2021integer}.
For such systems the flow reversal happens inside the core region.

\subsection{Comparison with numerical simulations}

\begin{figure*}[ht]
\begin{subfigure}{0.5\textwidth}\centering
  \includegraphics[width=\textwidth]{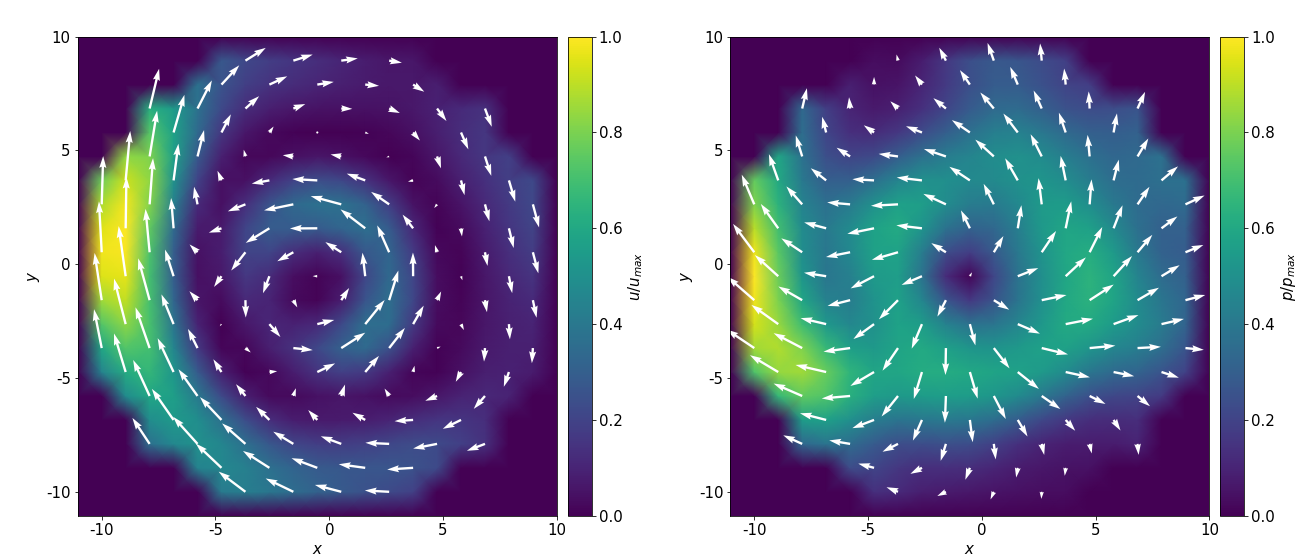}%
  \caption{$\tilde{\alpha}_0 = +0.4$}
  \label{subfig:-0.4+1def}
  \par 
  \medskip 
  \includegraphics[width=\textwidth]{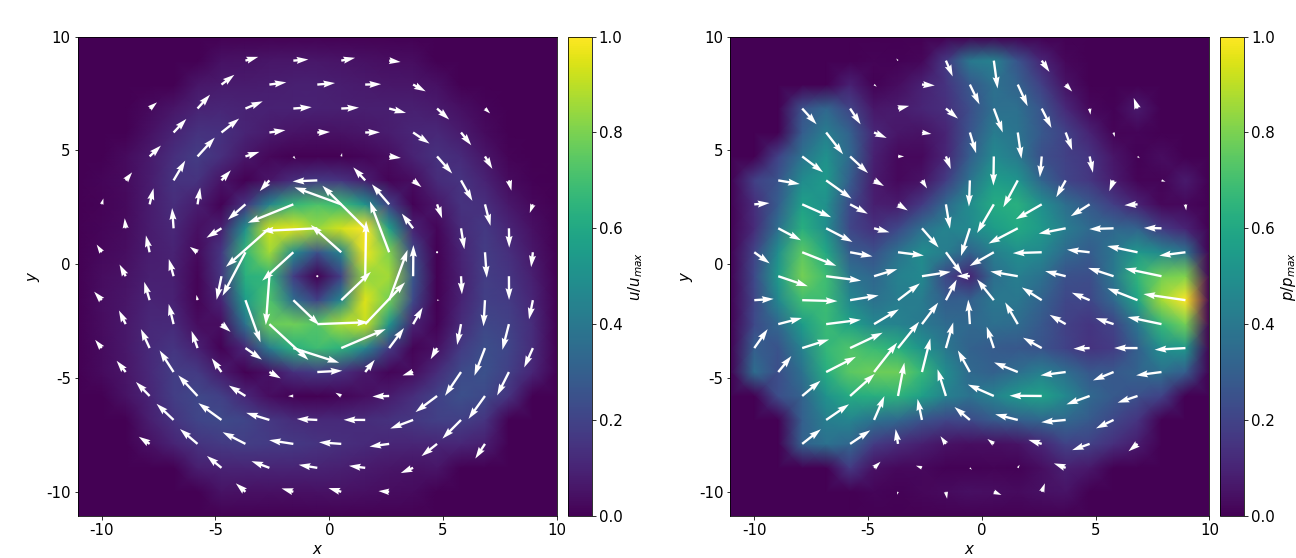}
  \caption{$\tilde{\alpha}_0 = +1$}
  \label{subfig:-1+1def}
\end{subfigure}
\hspace*{\fill}
\begin{subfigure}{0.5\textwidth}\centering
  \includegraphics[width=\textwidth]{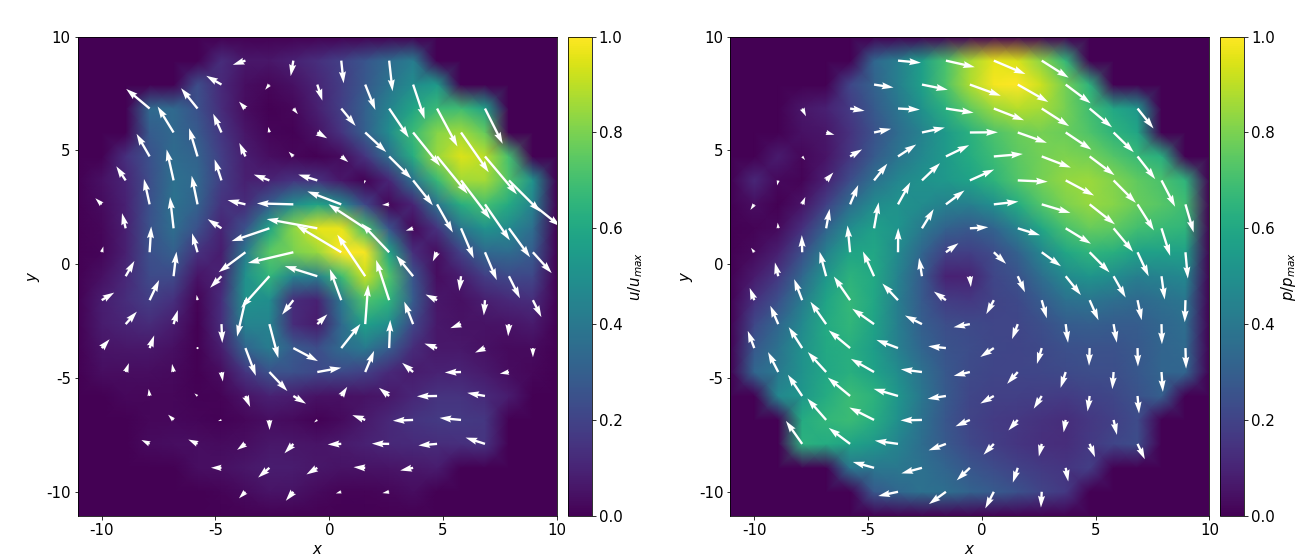}
  \caption{$\tilde{\alpha}_o= -0.4$}
  \label{subfig:+0.4+1def}
  \par 
  \medskip 
  \includegraphics[width=\textwidth]{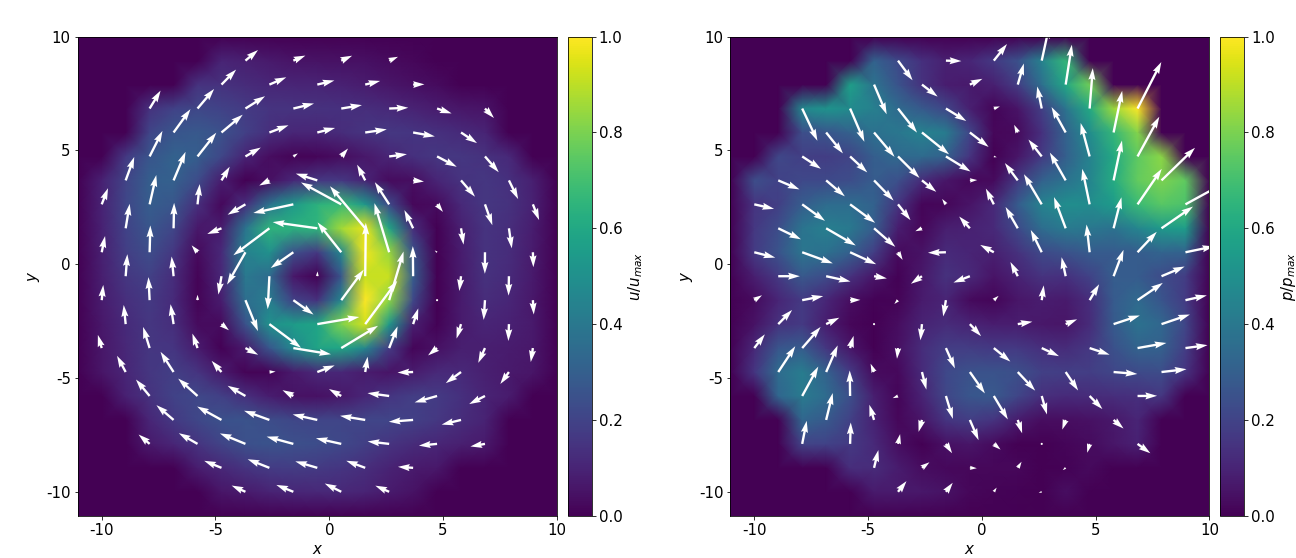}
  \caption{$\tilde{\alpha}_0 = -1$}
  \label{subfig:+1+1def}
\end{subfigure}
\caption{\textbf{Numerical average polarity and velocity fields around $(+1)-$defects, for various activities.} $\vec p$ (resp. $\vec u$) field is plotted on the right (resp. left) of each panel. Averages were computed on $\simeq 100$ defects for $\tilde \alpha_0 = \pm 0.4$, and on $\simeq 1000$ defects for $\tilde \alpha_0 = \pm 1$. The polarity field consists each time of a statistical combination of asters, spirals and vortices.
}
\label{fig:numerics_pos}
\end{figure*}

\begin{figure*}[t]
\begin{subfigure}{0.5\textwidth}\centering
  \includegraphics[width=\textwidth]{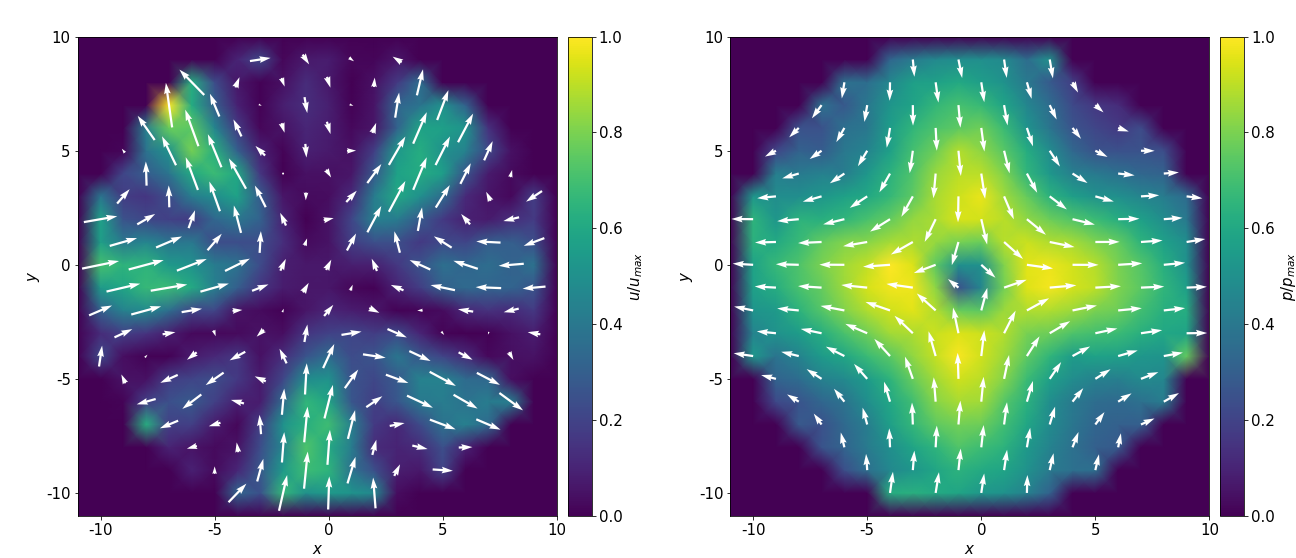}%
  \caption{$\tilde{\alpha}_0 = +0.4$}
  \label{subfig:-0.4-1def}
  \par 
  \medskip 
  \includegraphics[width=\textwidth]{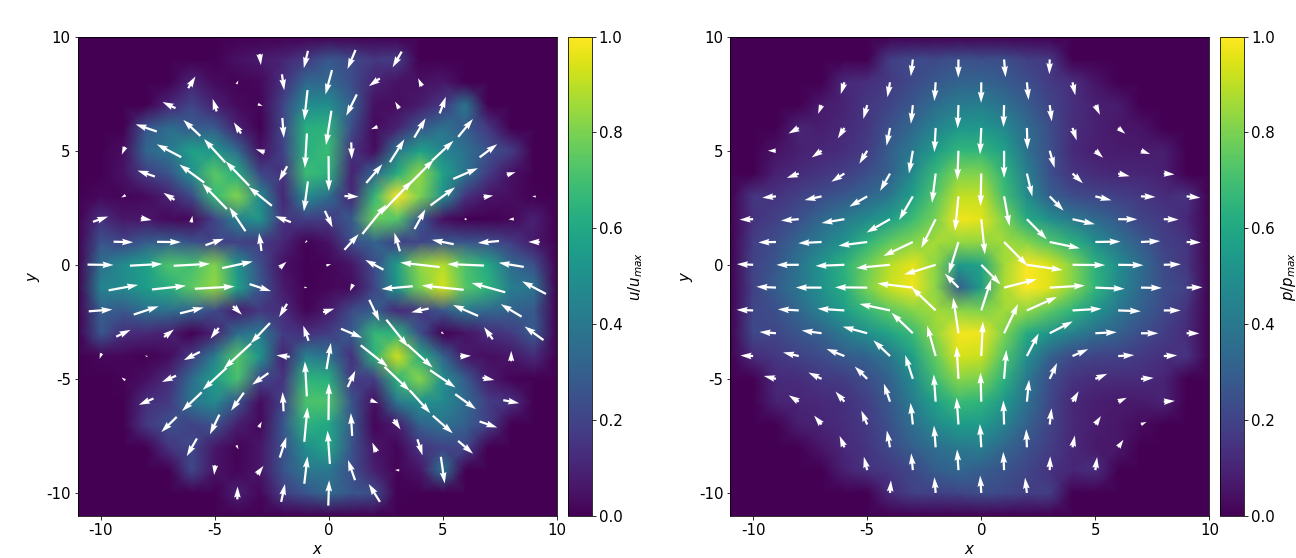}
  \caption{$\tilde{\alpha}_0 = +1$}
  \label{subfig:-1-1def}
\end{subfigure}
\hspace*{\fill}
\begin{subfigure}{0.5\textwidth}\centering
  \includegraphics[width=\textwidth]{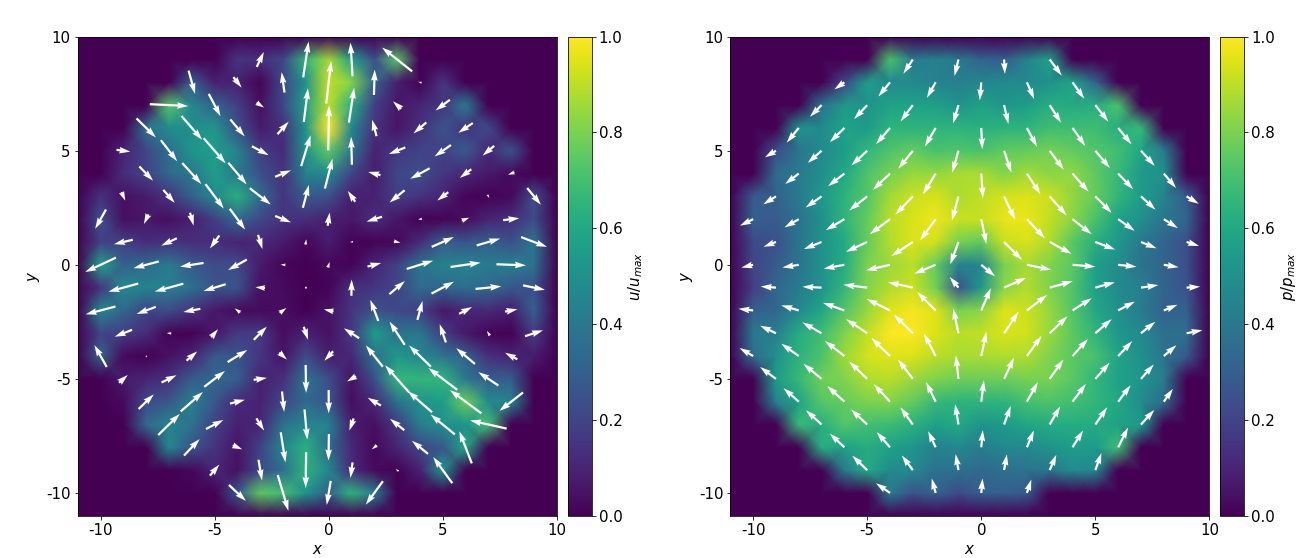}
  \caption{$\tilde{\alpha}_0 = -0.4$}
  \label{subfig:+0.4-1def}
  \par 
  \medskip 
  \includegraphics[width=\textwidth]{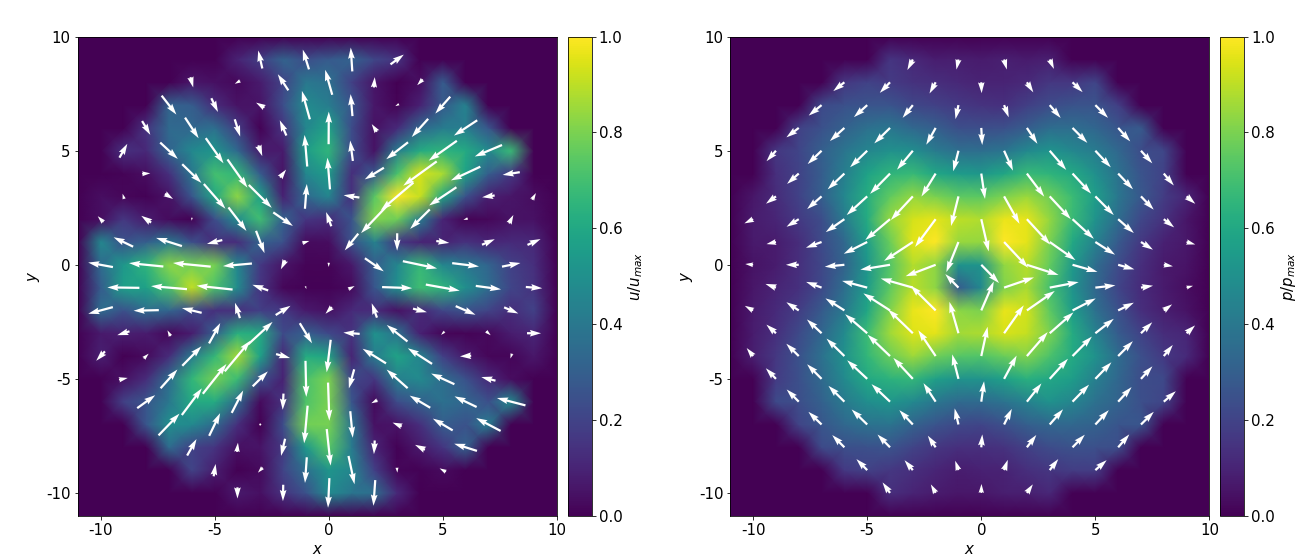}
  \caption{$\tilde{\alpha}_0 = -1$}
  \label{subfig:+1-1def}
\end{subfigure}
\caption{\textbf{Numerical average polarity and velocity fields around $(-1)-$defects, for various activities.} $\vec p$ (resp. $\vec u$) field is plotted on the right (resp. left) of each panel. Averages were computed on $\simeq 100$ defects for $\tilde \alpha_0 = \pm 0.4$, and on $\simeq 1000$ defects for $\tilde \alpha_0 = \pm 1$. Abnormal points (max. 8, on the edge of the averaging window) were set down to 0 to retain only significant flow patterns.
}
\label{fig:numerics_neg}
\end{figure*}

Here we discuss the comparison of the analytical predictions of active flow induced by $\pm 1$ defects with those obtained from direct numerical simulations of the polarization evolution Eq.~(\ref{eq:p}-\ref{eq:F}) coupled with the Navier-Stokes equations
\begin{equation}
    \rho(\partial_t \vec u +\vec u \cdot \nabla \vec u) =  \alpha_0 \nabla\cdot \mathbf{Q}-\nabla P +\nabla\cdot\mathbf{\sigma}_p, \quad \nabla\cdot\vec u =0
\end{equation}
with the additional passive stresses $\mathbf{\sigma}_p$ including the viscous stress as defined in Ref.~\cite{doostmohammadi2017onset}.
In numerical simulations, we use a hybrid lattice-Boltzmann method, combining finite-difference for the evolution of polarity and the lattice-Boltzmann method for that of velocity. Using the same prescription as in Ref.~\cite{andersen2022symmetry} for density and viscosity, i.e. $\rho=40$ and $\eta=3.6$, we ensure that the Reynolds number in the simulations remains negligible ($Re \ll 1$)~\cite{thampi2014instabilities,doostmohammadi2017onset} so that the dynamics of velocity virtually reduces to the incompressible Stokes equations Eq.~(\ref{eq:Stokes}-\ref{eq:incompressibility}) considered here. We fix the viscosity ratio to $\eta/\gamma=3.6$, micro to macro length scale to $(\sqrt{K_p/A})/L=2\times 10^{-3}$ (assuring that the coherence length $\sqrt{K_p/A}$ is significantly smaller than the domain size $L$), and the flow alignment parameter to $\lambda=0.1$.
Dimensionless parameters in the simulations are defined similar to the theoretical parameters defined in Section~\ref{sec2}.

Simulations were initialized with quiescent velocity field and noisy polar alignments close to the uniformly oriented state $\vec{p} = \vec{e}_x$ under periodic boundary conditions, on square domains of linear dimension $L=256$.

We ran simulations with sufficiently large values of the activity parameter $\tilde \alpha_0$ for $\pm 1$ topological defects to form spontaneously~\cite{andersen2022symmetry}. We compile the flow profiles in the vicinity of each defect for a large sample of defects (ranging from $\simeq 10^2-10^3$) for each simulation. From these individual flow profiles, we compute the statistically-averaged flows generated by $\pm 1$ topological defects.
However, because every defect had its own orientation, the flows had to be carefully reoriented upstream of the averaging procedure, as we detail below.

The orientation of $-1$ defects was defined from the surrounding polarity field. Given the 2-fold symmetry of the $-1$ defect (see fig.~\ref{fig:defect_streamlines}d)), we search for the two principal axes along which the polarity vector points inwards and outwards, respectively, and align the local fields accordingly.

Around $+1$ defects, both the polarity field and the velocity field are rotationally-symmetric (see Fig.~\ref{fig:defect_streamlines}a,b and Fig.~\ref{fig:vortex_stream_frictionless}). However, the vortical velocity around $+1$ defects may have different chirality: negative ("clockwise") when $u_{\theta}<0$, or positive ("counterclockwise") when $u_{\theta}>0$ as in Fig.~\ref{fig:vortex_stream_frictionless}. When simply summed up, these two chirality would cancel out, even though they have equivalent dynamic properties. To avoid this, we identify the flow chirality of each defect before averaging the fields. The flow chirality of each $+1$ defect was probed systematically by computing the sign of the circulation of $\vec u$ along small circular contours enclosing the defect center within the core of the defect, i.e. with the contour radius smaller than the length scale of the flow reversal $\sqrt{e}$. 
Defects with \textit{clockwise} chirality ($\oint\ u_{\theta} < 0$) are reversed ($p_{\theta} \leftarrow - p_{\theta}, u_{\theta} \leftarrow - u_{\theta}$), while those with \textit{counterclockwise} chirality are kept with the same orientation, so that all fields are consistently averaged. Note that no further distinction is made with regard to the polarity field, which therefore aggregates asters, spirals and vortices, since they lead to the same flow chirality.

The numerical results of the average polarity and flow profiles around $\pm 1$ defects are shown in Fig.~\ref{fig:numerics_pos} and Fig.~\ref{fig:numerics_neg}. First, for both types of defects, the profile of the polarity field (right panels in Fig.~\ref{fig:numerics_pos} and Fig.~\ref{fig:numerics_neg}) is consistent with the ideal vortex ansatz made by neglecting the backflow and the finite size of the core. The symmetries -- of the average -- are notably the same as in the passive case (see Fig.~\ref{fig:defect_streamlines}): positive defects are rotationally invariant -- seemingly with a mixture of asters, spirals and vortices --, while negative defects feature a 2-fold symmetry. Nonetheless, since our model allows the polarity vector to shrink and vanish, the local magnitude of polarity constitutes an additional degree of freedom that yields qualitatively robust patterns around $(-1)-$defects, yet with a shift by a $\pi/4-$rotation between contractile ($\bar \alpha_0 > 0 $) and extensile ($\bar \alpha_0 < 0 $) systems, as can be seen from the colored fonts in fig.~\ref{fig:numerics_neg}( \subref{subfig:-0.4-1def}, \subref{subfig:-1-1def} and \subref{subfig:+0.4-1def}, \subref{subfig:+1-1def} ). These patterns become smaller as the net activity increases since the active length-scale is inversely proportional to the activity $l^{\text{active}}\sim \sqrt{K/\alpha_0}$ (see fig.~\ref{fig:numerics_neg} \subref{subfig:-0.4-1def}, \subref{subfig:+0.4-1def} and \subref{subfig:-1-1def}, \subref{subfig:+1-1def} ). 

We now discuss the flow fields generated by negative defects. As predicted analytically above, the negative full-integer defects feature 8-fold rotational symmetry, with regions where the velocity points alternatively inwards and outwards, interspersed with counter-rotating swirls. This structure is relatively robust, and strengthens as the level of activity increases. The flow structure is also very similar for contractile and extensile systems, up to a rotation by the same angle $\pi/4-$ as above. In other words, equivalently, their profiles match, provided that one reverses the direction of the polarity vector according to the sign of the activity, which is consistent with the factor $\bar \alpha_0$ in the expression of the field $\vec u$ we previously derived.

Next, we discuss the velocity field generated by the positive defects. As expected from our theoretical considerations (fig.~\ref{fig:vortex_stream_pos}, fig.~\ref{fig:vortex_stream_frictionless}), the flow field exhibits a full-rotational symmetry. Since there was no friction, the dynamics was plainly viscous-dominated, and we have indeed striking evidence of the flow reversal predicted by our calculations in this regime, for any level of activity: as in the previous fig.~\ref{fig:vortex_stream_frictionless}, one can identify $2$ inner and outer regions rotating contrariwise in fig.~\ref{fig:numerics_pos}~\subref{subfig:-0.4+1def}-\subref{subfig:+1+1def}, that we will refer to as `core' and `shell' respectively. (In our captions, the core is rotating counterclockwise due to our averaging selection process). Taken together, the averaged velocity fields found from numerical simulation of polar active matter corroborate analytical predictions of the flow structures around full-integer defects and existence of counter-rotating vortices around positively charged defects. Furthermore, the simulation results were obtained from averaging the flow fields around full-integer defects within the active turbulence state and their close agreement with the analytical predictions around isolated defects give credit to the assumption of neglecting defect-defect interactions in the analytical calculations.

\subsection{Defect pair interaction modulated by dipolar active  forces}\label{Sec_3_3}
As seen in the above section the dipolar active force does not lead to any defect self-propulsion for isolated $\pm 1$ defects in an infinite domain. 
We now consider a pair of $\pm 1$ defects and show that the dipolar active force leads to an interaction between them through the active flow. 
For analytic tractability, we consider the friction-dominated regime $\Gamma \gg \eta$ and
set the viscosity to zero, $\ell_d \rightarrow 0$.
The defect velocity is calculated using the methods from \cite{angheluta2021role} (see \cite{SI} for details) and is given by 

\begin{align}
    u^+_a = \frac{\Tilde{\alpha}_0}{\Gamma R} e^{i(3\varphi -2\phi)}\left(\ln\left(\frac{R}{\Lambda}\right) -\frac{3}{2} \right), \\
    u^-_a = \frac{\Tilde{\alpha}_0}{\Gamma R} \left( \frac{1}{6}e^{i(3\varphi-2\phi)} +\frac{1}{4}e^{i(2\phi-\varphi) }\right),
    \end{align}
for the $+1$ defect and $-1$ defect, respectively. 
Here $\Lambda$ is a regularising lower cut-off corresponding to the finite defect core due to the divergent pressure at the $+1$ defect. $R$ is the distance between the defects, $\varphi$ is the angle of $\vec R = \vec r_- -\vec r_+$. $\phi$ is the uniform background orientation field. This effective interaction induced by dipolar active forces decays inversely proportional with the distance between defects similar to the Coulomb-like force induced by the phase gradients alone. However, these interactions are anisotropic since they depend on the orientation of the background polarization as well as the orientation of the defect pair.

\section{The effect of polar active forces} \label{sec4}
\begin{figure}
    \centering
     \includegraphics[width=0.3\textwidth]{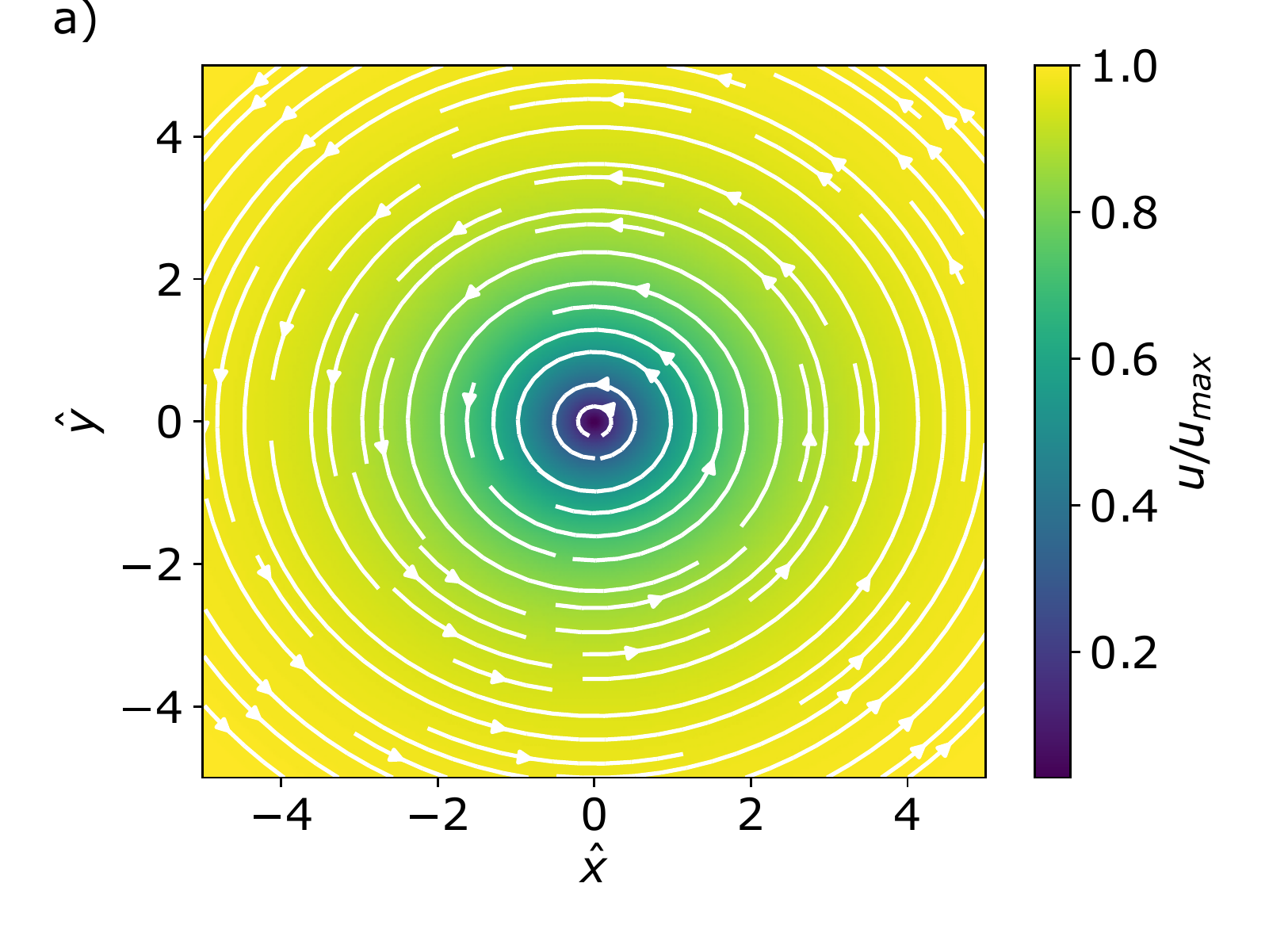}
    \includegraphics[width = 0.3\textwidth]{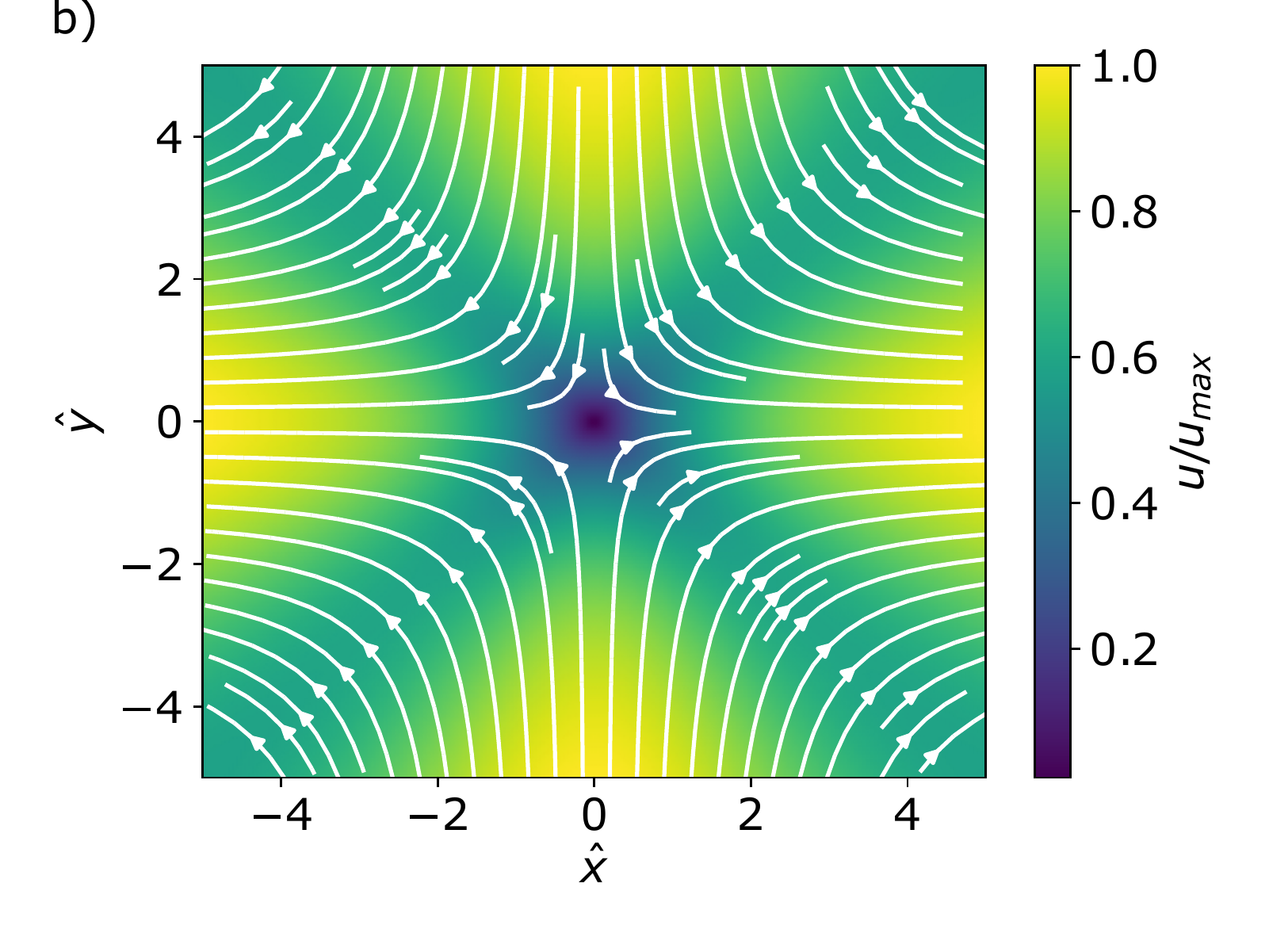}
    \caption{Velocity streamlines due to polar active forces for a) the positive defect (vortex flow) and b) the negative defect (saddle-point flow). Colormap represents the magnitude of the velocity field normalized by its maximum value. }
    \label{fig:polar_velocityfield}
\end{figure}

\begin{figure}[ht]
    \centering
    \includegraphics[width=0.22\textwidth]{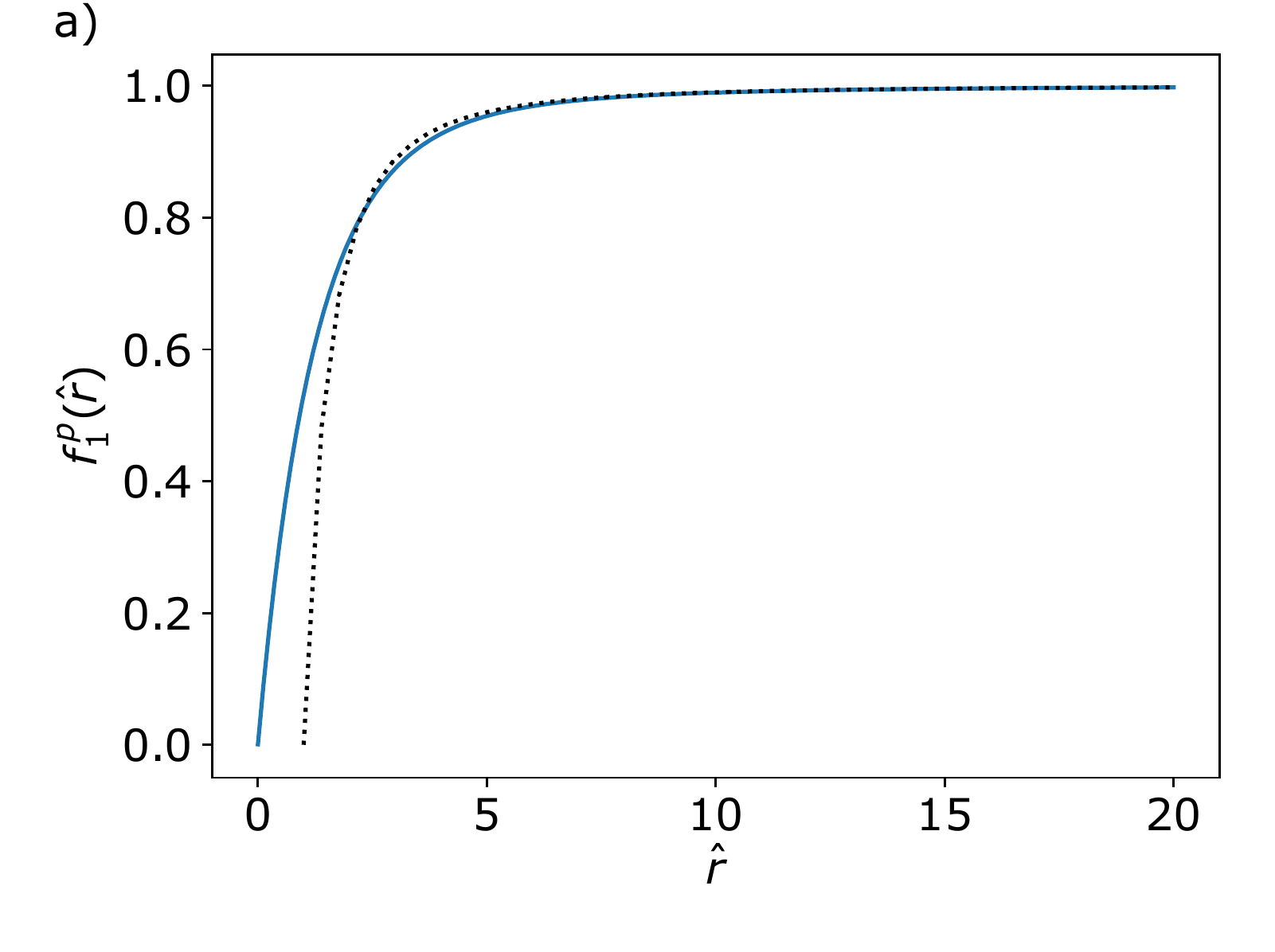}
    \includegraphics[width = 0.22\textwidth]{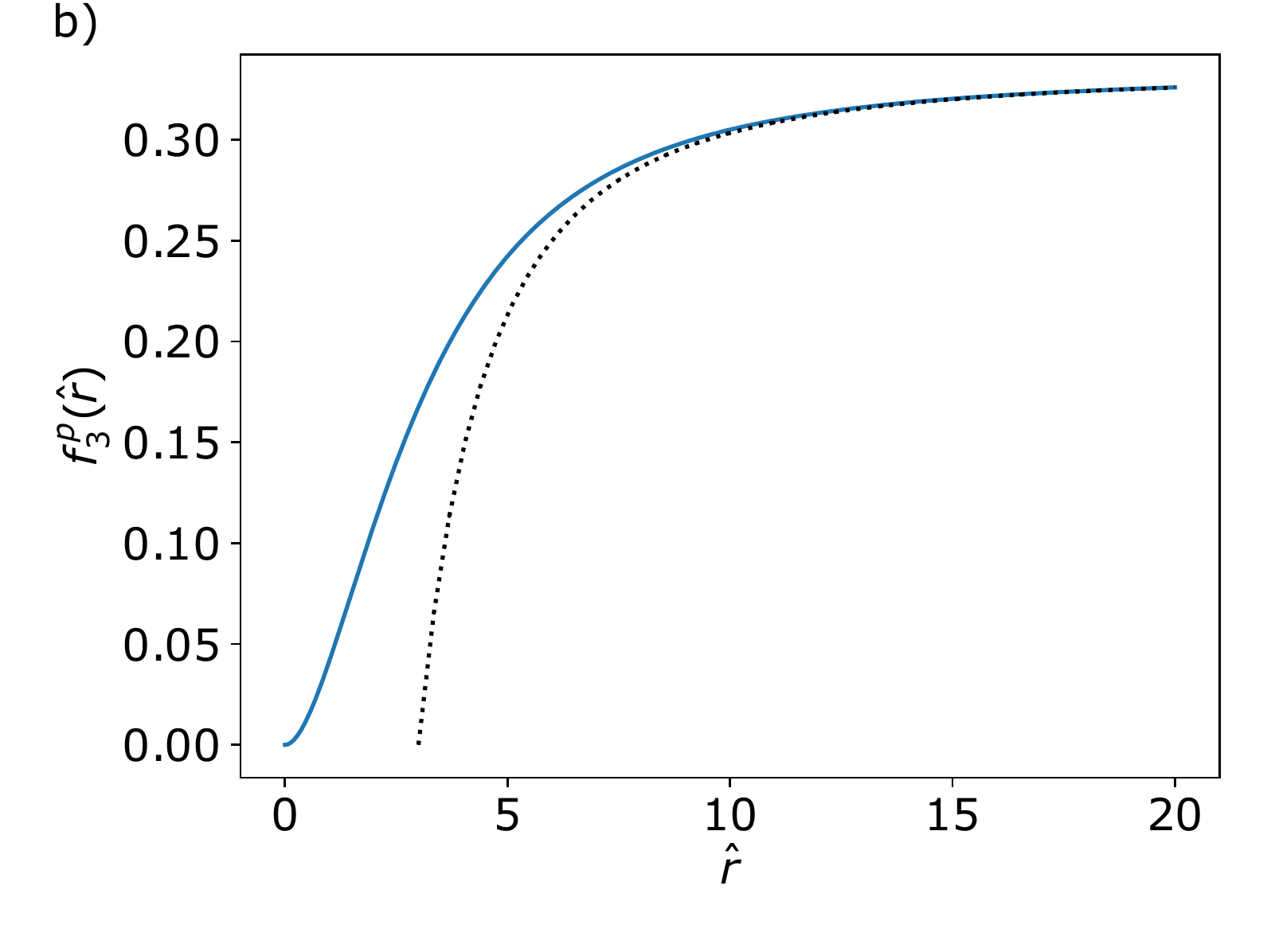}
    \caption{Plot of the functions (blue, whole lines) $f^p_1$ (a) and $f^p_3$ (b) against their asymptotic limits (black, doted lines) as given in eq.~(\ref{eq:Negative_polar_velocity_asymptot}) and (\ref{eq:Positive_polar_velocity_asymptot}). Note that the functions asymptotically approach to non-zero values in the far-field. }
    \label{fig:polar_functions}
\end{figure}

Next, we also include the polar active force in the main flow equations Eq.~(\ref{eq:flow_eqs}) and study its contribution to the flow velocity $\mathbf u_p$ and pressure $P_p$. In dimensionless units, the polar active force is 
\begin{equation}
    \vec F_p = \tilde{\alpha}_p \vec p.
\end{equation}
where $\tilde{\alpha}_p = \alpha_p\tau/(\xi\Gamma)$ is a rescaled parameter that measures the strength of polar active forces relative the frictional drag. We consider $\alpha_p > 0$ corresponding to polar particles moving in the direction of their head.
The corresponding flow velocity and pressure fields can now be written compactly as (see details in SM \cite{SI})
\begin{align}
    u_p^+(\hat r,\theta) &= i\Tilde \alpha_p e^{i\theta} \sin(\phi) f^p_1(\hat r),\\
    P_p^+( r,\theta) &= \Tilde \alpha_p r \cos(\phi) ,\label{eq:pplus}\\
    u_p^-(\hat r,\theta) &= \frac{\Tilde \alpha_p}{2}(f^p_1(\hat r)e^{-i\theta} +f^p_3(\hat r)e^{3i\theta} ), \\
    P_p^-(r,\theta) &= \frac{\Tilde \alpha_p }{3}\frac{(x^2 -y^2)}{r}. 
\end{align}
for the positive and the negative defect respectively. 
The velocity profiles are plotted in the Fig.~\ref{fig:polar_velocityfield} while the functions $f_1$ and $f_3$ are plotted in Fig.~\ref{fig:polar_functions} together with their asymptotic values. It is straightforward to show that both $f_1$ and $f_3$ tend to zero at the defect origin, thus the polar active forces do not contribute to defect self-induced motility. 
It is important to note that the flow induced by polar active forces around the $+1$ defect is proportional to $\sin(\phi)$, instead of the $\sin(2\phi)$ dependence that was found for the flow induced by dipolar active forces (Eq.~\ref{eq:pos_vel}). This means that while the polar active forces do not produce any flow around an ideal aster ($\phi=0$), there will be a polar activity-induced flow around an ideal vortex ($\phi = \pi/2$). 
Furthermore, Eq.~\ref{eq:pplus} shows that the pressure is proportional to the radial distance $\sim r$, indicating that, in the absence of any far-field screenings from other defects, the pressure can become quite large. This is related to the flow induced by polar active forces being constant in magnitude in the fare-field, as illustrated in Fig.~\ref{fig:polar_functions}, 
or by explicitly writing the asymptotic limits
\begin{align}
    u_p^-(\hat r,\theta) &=  \frac{\Tilde \alpha_p}{2}
    \Bigg[\left(e^{-i\theta} +\frac{1}{3} e^{3i\theta} \right)-\frac{1}{\hat r^2}\left[e^{-i\theta}+3e^{3i\theta}\right]  \Bigg],
    \label{eq:Negative_polar_velocity_asymptot} \\
      u_p^+(\hat r,\theta) &=  i\Tilde \alpha_p e^{i\theta} \sin{\phi}
   \left(1 -\frac{1}{\hat r^2} \right),
   \label{eq:Positive_polar_velocity_asymptot}
\end{align}
for the negative and positive defect respectively.  
As we will show explicitly for a pair of defects in the next section, Eq.~\ref{eq:Negative_polar_velocity_asymptot}-\ref{eq:Positive_polar_velocity_asymptot}, together with Eq.~(\ref{eq:point_vortex_model}), indicate that there are strong interactions between defects regardless of the distance between them.

Having found closed form formulas for the velocity field around the defects, the far-field vorticity induced by the positive defect can be calculated as
\begin{equation}
    \omega^+_p = \frac{\Tilde\alpha_p}{\zeta}\sin{\phi} \left(\frac{1}{\hat r} + \frac{1}{\hat r^3}\right),
\end{equation}
and, for the negative defect, this is
\begin{equation}
    \omega_p^- = \frac{2\Tilde\alpha_p }{\zeta}\sin{2\theta}\left(\frac{1}{\hat r} -\frac{3}{\hat r^3}\right).
\end{equation}
Interestingly the vorticity of the positive defect is independent on the polar angle $\theta$, which is also true in the near-field, because we can write the velocity field in the form $\vec u = \vec r^\perp f(r)$. As such, close to the defect center the vorticity of the negative defect vanishes at the center,
while the positive defects has a finite vorticity
\begin{equation}
    \omega_p^+(r=0) = \frac{\Tilde\alpha_p \pi}{2\zeta} \sin(\phi),
\end{equation}
indicating that the positive full-integer defect will be endowed with a spin.

\subsection{Defect-pair interaction in friction-dominated system}

Taking again the analytically tractable limit of zero viscosity, we can find the velocity induced by polar active forces for a pair of oppositely-charged defects following the approach from Sec.~\ref{Sec_3_3}. The flow velocities in the centre of the $\pm 1$ defects reduce to
\begin{align}
     \vec u^+_p &= \frac{\Tilde\alpha_p}{2 |\vec R|^2}  \vec R^\perp (\vec R^\perp \cdot \hat p_0), \label{eq:polar_dipole_pos} \\
     \vec u_p^- &= \frac{\Tilde\alpha_p}{6 |\vec R|^2}\vec R^\perp(\vec R^\perp \cdot \vec p_0)  + \frac{\alpha_p}{3|\vec R|^2} \vec R(\vec R \cdot \vec p_0), \label{eq:polar_dipole_neg}
\end{align}
where $\vec R  = \vec r^- -\vec r^+$ is the separation vector between the negative defect and the positive one, and $\vec p_0 = \cos(\phi) \vec e_x + \sin(\phi) \vec e_{-}$ is the uniform background polarity.
Interestingly, the polar active forces induce non-reciprocal and non-local mutual interactions that
depend only on the orientation of $\vec R$ relative to $\phi$, independent of the separation distance $R$ between the defects. This suggests a truly long-ranged interaction between oppositely-charged defects in the presence of polar active forces.

Furthermore, the $-1$ defect tends to move towards/away from $+1$ depending on the orientation angle of $\vec R$ with respect to $\vec p$. Both defects move perpendicular to $\vec R$ but at different rates inducing the pair to rotate. We can see this behavior more clearly by looking at the defect pair velocity under polar active forces alone, $\dot{\vec R} = \vec u_p^- - \vec u_p^+$ given by 
\begin{equation}
    \frac{d}{dt} \vec R=  \frac{\Tilde\alpha_p}{3 R^2} (\vec R \cdot \vec p_0)\vec R - \frac{\tilde \alpha_p}{3 R^2}(\vec R^\perp \cdot \vec p_0)\vec R^\perp,
\end{equation}
where the first term ($\parallel \vec R$) is an attraction/repulsion force between the defects, while the second term ($\perp\vec R$) rotates the defect pair.  
The rotation is zero, when the defect pair aligns with the background polarization in either directions ($\vec R \parallel \vec p_0$). We show in SM.~\cite{SI} that the defect pair rotates until annihilation, unless $\vec R$ is initially parallel to $\vec p_0$. This shows that the polar active force leads to pair annihilation of oppositely-charged defects to promote large-scale polar order. 

To check this interesting dynamics numerically, we compare the pair trajectory 
determined by Eq.~(\ref{eq:point_vortex_model}) with that predicted by the full hydrodynamic model from integrating Eqs.~(\ref{eq:p}) (\ref{eq:Stokes}) and (\ref{eq:incompressibility}) with the same initial configuration. The initial uniform polarisation $\vec p = \vec e_y$ is seeded with a defect pair in the $x$-direction, i.e. $\vec r_- =-20\vec e_y$ and  $\vec r_+ = 20 \vec e_x$, such that $\vec R\perp \vec p_0$ at $t=0$. The model parameters are set to $\tilde \alpha_p =0.5$, $\tilde \alpha_0 =0$, $\lambda=0$ and $\zeta^2 =0.01$. The hydrodynamic equations are solved with periodic boundary conditions in a domain $256\times256$ with a spatial discretisation $\Delta=0.5$, using spectral methods and an exponential time differentiation scheme~\cite{cox2002exponential}. 
Under the polar active force, the separation vector $\vec R$ rotates relative to $\vec p_0$, and this changes the shape of the $+1$ defect from an initial vortex to an aster at the annihilation time. This effect is lost in the absence of polar active forces, i.e.at $\tilde\alpha_p=0$ (see animations of the defect pair annihilation in~\cite{SI}). 

\begin{figure}[ht]
\centering
  \includegraphics[width = .3\textwidth]{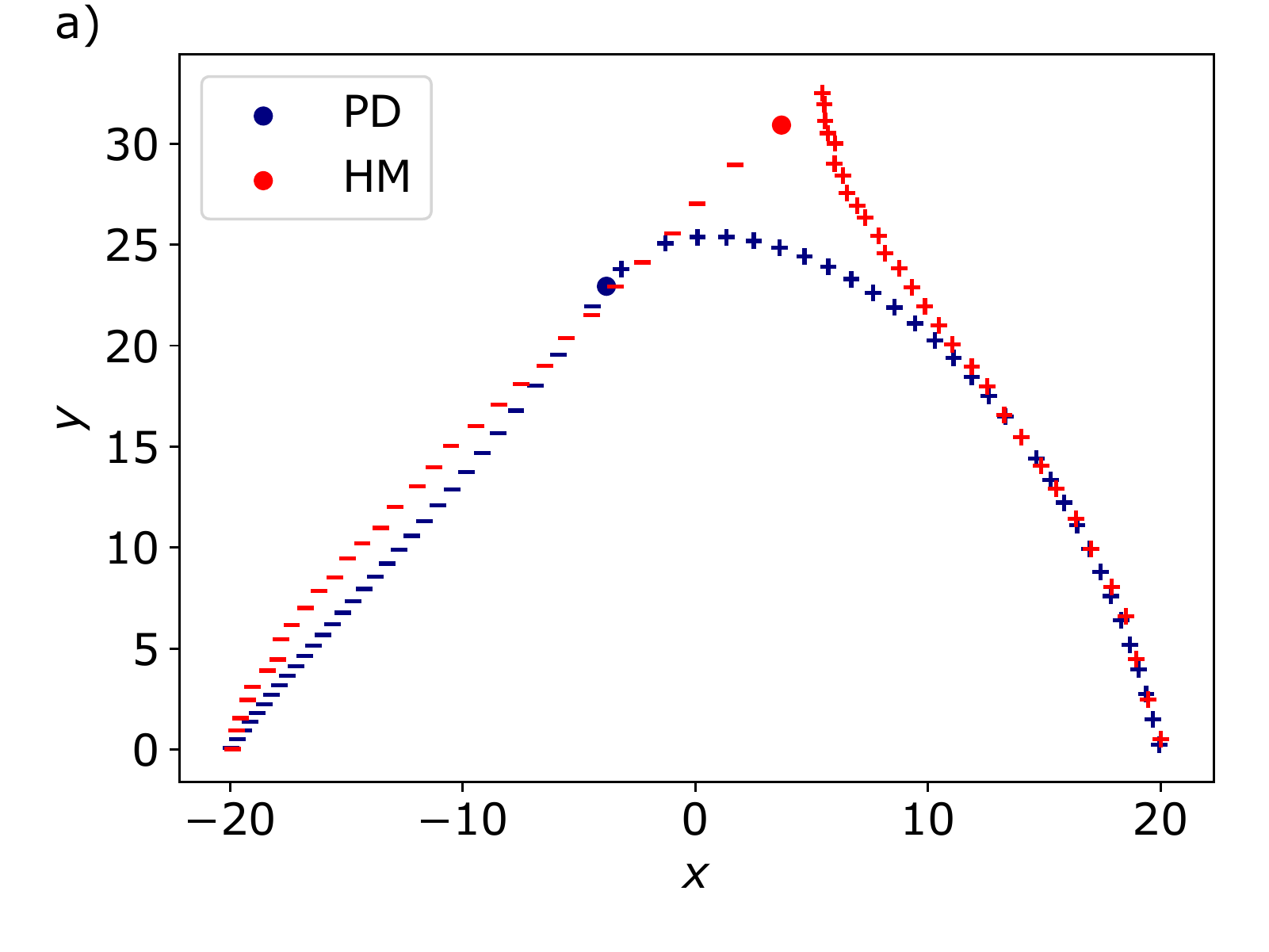}
  \includegraphics[width = .3\textwidth]{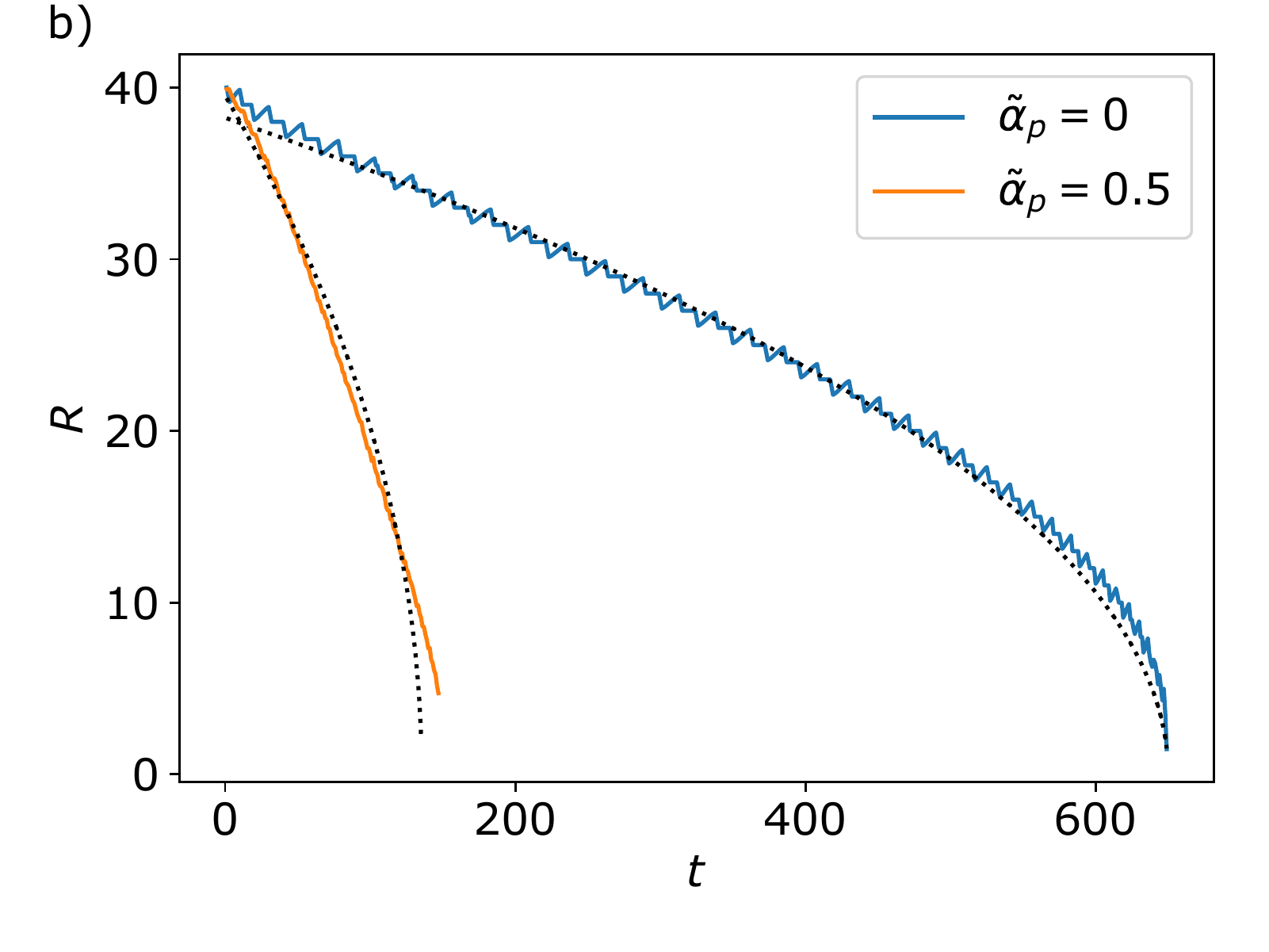}
\caption{a) Phase portraits of the defect pair trajectory obtained from (blue) the kinematic law for point defects (PD) from Eq.~(\ref{eq:point_vortex_model}), and (red) by direct simulation of the full hydrodynamic model (HM) starting from the same initial condition. The positive defect is marked by $+$, the negative by $-$. $\tilde \alpha_p =0.5$, $\tilde \alpha_0 =0$, $\lambda=0$ and $\zeta^2 =0.01$. b) Distance $R$ between a defect pair as a function of time for Coulomb-like interactions only (blue curve) and in the presence of polar active forces (orange curve). The black doted lines are fitting curves with the $R(t) = \sqrt{At +B}$ found by linear regression on $R^2$.} \label{fig:defect_trajectories}
\end{figure}

The two trajectories are shown in Fig.~\ref{fig:defect_trajectories}a) and agree very well for large $R= |\vec R|$ in consistency with the pointwise approximation. It also shows that polar active forces increase the annihilation rate. This is further evidenced in Fig.~\ref{fig:defect_trajectories} b) where we compare the evolution of the separation distance $R(t)$ when the two defects interact through the Coulomb-like forces with or without the presence of polar active forces. At $\tilde\alpha_p=0$, $R(t)$ decreases in the far-field as $\sqrt{t_0-t}$ ($t_0$ being the annihilation time) due to the $1/R$ interaction forces. However, for $\tilde\alpha_p\ne 0$, the annihilation timescale is greatly reduced and the deviations from the $\sqrt{t_0-t}$ behavior arise due the non-local and non-reciprocal attraction force. The rotation rate is also contributing to aligning the defects to increase the attraction between the defects.

\section{Discussion/conclusion}

In summary, we present theoretical derivations of incompressible flow fields and dynamics of $\pm 1$ defects under dipolar and polar active forces. These active forces do not endow the full-integer defects with any self-propulsion, as expected from the symmetry of the defects. However, the $+1$ defect acquires a non-zero active torque due to both polar and dipolar active forces. We show that the strength of this spin is dependent on whether the defect is a vortex, aster or spiral.
In the absence of hydrodynamic screening due to friction, the vortical active flow around a $+1$ vortex changes sign on a length scale set by the coherence length as observed in numerical simulations and predicted analytically. For both defects, the dipolar active force contribution to the flow field vanishes with $1/r$ in the defect far-field, whereas the polar active force contribution approach a constant value. Remarkably, polar active forces mediate mutual interaction between oppositely-charged defect pairs in a manner that renders defect-defect interactions independent of the distance between the defect pair. We have shown that, under these polar active forces, a pair of oppositely-charged defects rotates to align with the background polarization field, while the negative defect chases the positive one until annihilation. The rate of annihilation is greatly enhanced by polar active forces, and this is the main underlying mechanism for the suppression of defect-laden active turbulence in polar active matter as reported numerically in Ref.~\cite{andersen2022symmetry}. 
\section*{Author Contributions}
J.R{\o} contributed with analytical derivations, numerical simulations, visualisation and data analysis. J.Re. contributed with lattice-Boltzmann simulations, visualisation and data analysis. A.D. and L.A. formulated the project and contributed with supervision. A.D. provided computational resources. L.A. checked the analytical calculations. All authors contributed with drafting and writing the manuscript.  

\section*{Conflicts of interest}
There are no conflicts to declare.

\section*{Acknowledgements}
 J.R{\o}. and L.A. acknowledge support from the Research Council of Norway through the Center of Excellence funding scheme, Project No. 262644 (PoreLab). A. D. acknowledges funding from the Novo Nordisk Foundation (grant No. NNF18SA0035142 and NERD grant No. NNF21OC0068687), Villum Fonden Grant no. 29476, and the European Union via the ERC-Starting Grant PhysCoMeT. Views and opinions expressed are however those of the authors only and do not necessarily reflect those of the European Union or the European Research Council. Neither the European Union nor the granting authority can be held responsible for them.



\balance


\onecolumn
\section*{Appendix: }
Here we write down the radially-dependent functions associated with the flow fields induced by polar and dipolar active forces. Derivation details are found in the SM.~\cite{SI}.
The radial functions for the dipolar active flow velocity are:
\begin{align}
    f_3^a(\hat r) &= \frac{1}{\hat r}[1-\hat r K_1(\hat r)]- \frac{2}{\hat r^3}[4-\hat r^3 K_1(\hat r) - 2\hat r^2 K_2(\hat r)], \\
    f_5^a(\hat r) &= \frac{1}{\hat r} [1-\hat r K_1(\hat r)] -\frac{6 }{\hat r^3} [4-\hat r^3 K_1(\hat r) - 2\hat r^2 K_2(\hat r)]
    + \frac{6}{\hat r^5} \left[64-4\hat r^2( 8+\hat r^2)K_0(\hat r) - \hat r (8+\hat r^2)^2 K_1(\hat r) \right]. 
\end{align}
The radial function for the corresponding vorticity field 
\begin{equation}
     f_\omega^a(\hat r)=\frac{1}{\hat r^4}\left(-48 +4\hat r^2 +\hat r^2(24+\hat r^2)K_0(\hat r) +8\hat r(6+\hat r^2)K_1(\hat r) \right).
\end{equation}
The radial functions for the polar active flow velocity are:
\begin{equation}
    f^p_{1}( \hat r) =1 - \hat rK_1( \hat r) +\frac{\pi}{2} I_1(\hat r) + \sum_{k,n=0}^\infty \kappa_1(n,k) \frac{ \hat r^{2k+2}}{((2k)!!)^2} 
    + \sum_{k,n=0}^\infty \kappa_2(n,k) \left[\ln\left(\frac{\hat r}{2}\right)-\psi^{(0)}(k+1)\right] \frac{\hat r^{2k+2}}{((2k)!!)^2},
\end{equation}
and
\begin{equation}
    f^p_3(\hat r) = -\frac{\pi}{2}I_3(\hat r) + \sum_{k,n=0}^\infty \kappa_3(n,k) \frac{\hat r^{2k+2}}{((2k)!!)^2} 
    + \sum_{k,n=0}^\infty \kappa_4(n,k) \frac{\hat r^{2k+2}}{((2k)!!)^2}\left[\ln\left(\frac{\hat r}{2}\right) - \psi^{(0)}(k+1)\right].
\end{equation}
Where $\psi^{(0)}$ is the digamma function. 
The coefficients of the infinite power series
\begin{align}
    \kappa_1(n,k) &= - \left(\frac{(2n-1)!!}{(2n)!!}\right)^2\frac{1}{2n+2} \left(\frac{1}{(2n-1-2k)^2} +\frac{1}{4}\frac{2n+1}{(2n+2)(n+2+k)^2}\right) \\
    \kappa_2(n,k) &= \left(\frac{(2n-1)!!}{(2n)!!}\right)^2 \frac{1}{2n+2} \left( \frac{2n+1}{2 (2n+2) (n+2+k)} -\frac{1}{2n-1-2k}\right), \\
    \kappa_3(n,k) &= \left(\frac{(2n-1)!!}{(2n)!!}\right)^2 \left(\frac{2n+1}{4(3-2n)(1-2n)(n+k+1)^2} +\frac{2n}{(2n+4)(2n+2)(2n-1-2k)^2} \right), \\
    \kappa_4(n,k) &= \left(\frac{(2n-1)!!}{(2n)!!}\right)^2 \left( \frac{2n}{(2n+4)(2n+2)(2n-1-2k)} -\frac{2n+1}{2(3-2n)(1-2n)(n+k+1)}\right).
\end{align}

Note that when summing over $n$ the coefficients $\kappa_4(n,k)$ tends to zero for any $k$. The infinite series are convergent, but slowly.   

\twocolumn

\bibliography{references}
\bibliographystyle{rsc} 

\end{document}